\documentclass[12pt]{article}
 \usepackage{ifpdf}
 \usepackage{epsfig}
  \usepackage[latin1]{inputenc}
 \usepackage{amsfonts}
 \usepackage{amsthm}
 \usepackage{amssymb, latexsym, amsmath}
 \usepackage{amsbsy}
\usepackage{amstext}
\usepackage{amsopn}
\usepackage{amsgen}
\usepackage[dvips]{color}
 \usepackage{graphicx}
  \usepackage{color}
 \newcommand{\inc}{{\it i}}
 \newcommand{\etadot}{\dot{\eta}}
 \newcommand{\betadot}{\dot{\beta}}
 \newcommand{\etadotdot}{\ddot{\eta}}
 \newcommand{\thetadot}{\dot{\theta}}
 \newcommand{\thetadotdot}{\ddot{\theta}}

 \newcommand{\be}{\begin{equation}}
 \newcommand{\ee}{\end{equation}}
 \newcommand{\ba}{\begin{eqnarray}}
 \newcommand{\ea}{\end{eqnarray}}

 \newcommand{\erbold}{\mbox{{\boldmath $\vec r$}}}

 \newcommand{\taubold}{\mbox{{\boldmath $\vec \tau$}}}

\begin{document}
 \title{
                 ${{~~~~~~~}^{^{^{
                 ``Celestial~Mechanics~and~Dynamical~Astronomy"
          \,,~Vol.~114\,,~pp.~387\,-\,414\,~(2012).
                  }}}}$ ~\\
 {\Large{\textbf{Bodily tides near the $\,1:1\,$ spin-orbit resonance.\\
 Correction to Goldreich's dynamical model
 \\}
            }}}
 \author{
     {\Large{James G. Williams}}\\
     {\small{Jet Propulsion Laboratory, California Institute of Technology, Pasadena CA 91109 USA}}\\
     {\small{e-mail: ~james.g.williams @ jpl.nasa.gov~}}
  ~\\
  ~\\
  and\\
  ~\\
  {\Large{Michael Efroimsky}}\\
  {\small{US Naval Observatory, Washington DC 20392 USA}}\\
  {\small{e-mail: ~michael.efroimsky @ usno.navy.mil~}}
  }
 \date{}

 \maketitle
 \begin{abstract}

 Spin-orbit coupling is often described in an approach known as {\emph{``the MacDonald torque"}}, which has long become the
 textbook standard due to its apparent simplicity. Within this method, a concise expression for the additional tidal potential, derived by MacDonald (1964; {\it{Rev. Geophys.}} {\bf{2}}, 467 - 541), is combined with a convenient assumption that the quality
 factor $\,Q\,$ is frequency-independent (or, equivalently, that the geometric lag angle is constant in time). This makes the treatment unphysical because MacDonald's derivation of the said formula was, very implicitly, based on keeping the {\it{time}} lag frequency-independent, which is equivalent to setting $\,Q\,$ to scale as the inverse tidal frequency. This contradiction requires the entire MacDonald treatment of both non-resonant and resonant rotation to be rewritten.

 The non-resonant case was reconsidered by Efroimsky \& Williams (2009; {\it{Cel.Mech.$\,\&$ Dyn.Astr.}}
 {\bf{104}}, 257 - 289), in application to spin modes distant from the major commensurabilities. In the current paper, we continue
 this work by introducing the necessary alterations into the MacDonald-torque-based model of falling into a 1-to-1 resonance. (The
 original version of this model was offered by Goldreich 1966; {\it{AJ}}~ {\bf{71}}, 1 - 7.)

 Although the MacDonald torque, both in its original formulation and in its corrected version, is incompatible with realistic
 rheologies of minerals and mantles, it remains a useful toy model, which enables one to obtain, in some situations, qualitatively
 meaningful results without resorting to the more rigorous (and complicated) theory of Darwin and Kaula.

 We first address this simplified model in application to an oblate primary body, with tides raised on it by an orbiting
 zero-inclination secondary. (Here the role of the tidally-perturbed primary can be played by a satellite, the perturbing secondary
 being its host planet. A planet may as well be the perturbed primary, its host star acting as the tide-raising secondary.) We then
 extend the model to a triaxial primary body experiencing both a tidal and a permanent-figure torque exerted by an orbiting secondary.
 We consider the effect of the triaxiality on both circulating and librating rotation near the synchronous state. Circulating rotation
 may evolve toward the libration region or toward a spin faster than synchronous (the so-called pseudosynchronous spin). Which behaviour
 depends on the orbit eccentricity, the triaxial figure of the primary, and the mass ratio of the secondary and primary bodies. The spin
 evolution will always stall for the oblate case. For libration with a small amplitude, expressions are derived for the libration
 frequency, damping rate, and average orientation.

 Importantly, the stability of pseudosynchronous spin hinges upon the dissipation model employed. Makarov and Efroimsky (2013; arXiv:1209.1616)
 have found that a more realistic tidal dissipation model than the corrected MacDonald torque makes pseudosynchronous spin unstable. Besides,
 for a sufficiently large triaxiality, pseudosynchronism is impossible, no matter what dissipation model is used.

 \end{abstract}

 \section{Motivation}

 Bodily tides in a near-spherical homogeneous primary perturbed by a point-like secondary are described by the theory developed mainly
 by Darwin (1879, 1880) and Kaula (1966). Sometimes this theory is referred to as the {\it{Darwin torque}}. Based on a Fourier-like
 expansions of the perturbing potential and of the tidally-induced potential of the disturbed primary, their theory permits an
 arbitrary rheology of the primary.

 Sometimes a much simpler empirical model, offered by
 MacDonald (1964) and often named as the {\it{MacDonald
 torque}}, is used in the literature for obtaining very approximate but still qualitatively reasonable description of tidal evolution.
 This model can be derived from the Darwin-Kaula theory, under several simplifying assumptions. Among these is the assumption that the
 tidal quality factor $\,Q\,$ of the perturbed primary should scale as the inverse of the tidal frequency. Being incompatible with the
 rheologies of actual minerals, this key assumption prohibits the use of the MacDonald model in long-term orbital calculations.

 Despite the unrealistic rheology instilled into the MacDonald model, the model remains a lab which permits one to
 gain qualitative understanding of tidal evolution (rotational and orbital), without resorting to the lengthy calculations required in the accurate Darwin-Kaula approach. It should be noted however that employment of the MacDonald model needs some care. Historically, the orbital calculations performed with aid of this model by one of its creators, MacDonald (1964), contained an inherent contradiction. MacDonald began his paper with deriving a concise expression for the additional tidal potential, and then combined this expression with a convenient assumption that the geometric lag angle is constant (or, equivalently, that the tidal quality factor is a frequency independent constant). This made the treatment unphysical because MacDonald's derivation of the said formula was, very implicitly, based on keeping the {\it{time}} lag frequency-independent, which is equivalent to setting $\,Q\,$ to scale as the inverse tidal frequency. This contradiction made his theory inconsistent. The said oversight has then been repeated
 many a time in the literature (Kaula 1968, eqn. 4.5.37; Murray \& Dermott 1999, eqn. 5.14). \footnote{~Due to an error in our
 translation from German, we mis-assumed in our previous papers Efroimsky \& Williams (2009) and Efroimsky (2012a) that Gerstenkorn
 (1955) had based his development on a constant-$Q$ model. Therefore we stated that his theory contained the same
 genuine inconsistency as the theory by MacDonald (1964). Accurate translation of the work by Gerstenkorn (1955) has shown that his
 method was based on a constant-time-lag model. Therefore we retract our statement about Gerstenkorn's approach sharing the
 inconsistency of MacDonald's theory. We also thank Hauke Hussmann and Peter Noerdlinger for their kind help in translating excerpts
 from Gerstenkorn's work.} In particular, the MacDonald method was
 used by Goldreich (1966) in his theory of dynamical evolution near the 1:1 spin-orbit resonance. We reconsider this theory, employing the
 corrected version of the MacDonald torque, i.e., setting the quality factor to scale as inverse frequency.

 \section{Linear bodily tide in a near-spherical primary}


 Consider a near-spherical primary of radius $\,R\,$ and a secondary of mass $M^*_{sec}$ located at ${\erbold}^{\;*} = (r^*,\,\phi^*,\,
 \lambda^*)$, where $r^*\geq R$. The tidal potential created by the secondary alters the primary's shape and, as a result, its potential.
 For linear tides, the amendment to the primary's exterior potential is known (e.g., Efroimsky \& Williams 2009, Efroimsky 2012a,b) to
 read as
 \ba
 U(\erbold)\;=\;\,-\,{G\;M^*_{sec}}
 \sum_{{\it{l}}=2}^{\infty}k_{\it l}\;
 \frac{R^{
 \textstyle{^{2\it{l}+1}}}}{r^{
 \textstyle{^{\it{l}+1}}}{r^{\;*}}^{
 \textstyle{^{\it{l}+1}}}}\sum_{m=0}^{\it l}\frac{({\it l} - m)!
 }{({\it l} + m)!}(2-\delta_{0m})P_{{\it{l}}m}(\sin\phi)P_{{
 \it{l}}m}(\sin\phi^*)\;\cos m(\lambda-\lambda^*)\,~,~\quad
 \label{4}
 \ea
 $\delta_{ij}\,$ being the Kronecker delta, $\,G=6.7\times10^{-11}\,
 \mbox{m}^3\,\mbox{kg}^{-1}\mbox{s}^{-2}\,$ being Newton's gravity constant, and $\,\gamma\,$ being the angle between the vectors
 $\,\erbold^{\;*}\,$ and $\,\erbold\,$ pointing from the primary's centre. As agreed above, $\,{\erbold}^{\;*}\,=\,(r^*\,,\,\phi^*\,,\,
 \lambda^*)$ denotes the position of the perturber, while $\,\erbold\,=\,(r\,,\,\phi\,,\,\lambda)\,$ is an exterior point, at a radius
 $\,r\,\geq\,R\,$, where the tidal potential amendment $\,U(\erbold)\,$ is measured. The longitudes $\lambda,\,\lambda^*$ are reckoned
 from a fixed meridian on the primary, while the latitudes $\phi,\,\phi^*$ are reckoned from its equator.
 Indices $\,l\,$ and $\,m\,$ are traditionally referred to as the {\it{degree}} and {\it{order}}, accordingly. The associated Legendre
 functions $\,P_{lm}(x)\,$ (termed associated Legendre {\it{polynomials}} when their argument is sine or cosine of some angle) are
 introduced as in Kaula (1966, 1968), and may be called {\it{unnormalised}} associated Legendre functions, to distinguish them from their
 normalised counterparts.\footnote{~The Legendre polynomials may be defined through the Rodriguez formula
 $~
 P_l(x)
 \,=\,\frac{\textstyle 1}{\textstyle 2^l\,l!}~\frac{\textstyle d^{\,l}}{\textstyle dx^{\,l}}
 \left(\,x^2\,-\,1\,\right)^l~.~
 ~$
 In most literature, the associated Legendre functions are introduced, for a nonnegative $\,m\,$, as
 \ba
 P_{lm}(x)~=~\left(\,1\,-\,x^2\,\right)^{m/2}~\frac{d^m~}{dx^m}\,P_l(x)~\quad~\quad~\mbox{and}~\quad\quad
 P_{l}^{m}(x)~=~(-1)^{m}\,\left(\,1\,-\,x^2\,\right)^{m/2}~\frac{d^m~}{dx^m}\,P_l(x)~~~,
 \nonumber
 \ea
 so that
 $~
 P_{lm}(x)~=~(-1)^{m}\,P_{l}^{m}(x)~.
 ~$
 The above definition agrees with the one offered by Abramowitz \& Stegun (1972, p. 332). A different convention is accepted in those
 books (e.g., Arfken \& Weber 1995, p. 623) where $\,P_l^m(x)\,$ lacks the $\,(-1)^{m}\,$ factor and thus coincides with
 $\,P_{lm}(x)\,$.}
 The Love numbers $\,k_{\it l}~$ can be calculated from the geophysical properties of the primary.

 A different formula for the tidal-response-generated change in the potential was suggested by Kaula (1961, 1964). Kaula devised a
 method of switching variables, from the spherical coordinates to the orbital elements $\,(\,a^*,\,e^*,\,\inc^*,\,\Omega^*,\,\omega^*,\,
 {\cal M}^*\,)\,$ and $\,(\,a,\,e,\,\inc,\,\Omega,\,\omega,\,{\cal M}\,)\,$ of the secondaries located at $\,\erbold^{\;*}\,$ and
 $\,\erbold\,$. The goal was to explore how a tide-raising secondary at $\,\erbold^{~*}\,$ acts on a secondary at $\,\erbold\,$ through
 the medium of the tides it creates on their mutual primary.

 The development enabled Kaula to process (\ref{4}) into a series, which was a disguised form of a Fourier expansion of the
 tide. Interestingly, Kaula himself never referred to that expansion as a Fourier series, nor did he ever write down explicitly the
 expressions for the Fourier modes. At the same time, the way in which Kaula introduced the phase lags indicates that he was aware of
 how the modes were expressed via the perturber's orbital elements and the primary's rotation rate. The original works by Kaula (1961,
 1964) were written in a terse manner, with many technicalities omitted. A comprehensive elucidation of his approach can be found in the
 Efroimsky \& Makarov (2013). Referring the reader to that paper for details, here we cite only the resulting
 formula for the secular part of $\,U\,$, in the special case when the tide-raising secondary itself experiences perturbation from the
 tide it creates on the primary (so $\,\erbold^{~*}=\erbold\,$):
 \ba
 U^{(sec)}(\erbold)\,=~\quad~\quad~\quad~\quad~\quad~\quad~\quad~\quad~\quad~\quad~\quad~\quad~\quad~\quad~\quad~\quad~\quad~\quad
 ~\quad~\quad~\quad~\quad~\quad~\quad~\quad~\quad~\quad~\quad~\quad~\quad~\quad~\quad~\quad~\quad~\quad~\quad
 \nonumber
 \ea
 \ba
 \,-~\frac{G\,M_{sec}}{a}\,\sum_{l=2}^{\infty}\,\left(\,\frac{R}{a}\,\right)^{\textstyle{^{2l+1}}}
 \sum_{m=0}^{\it l}\frac{({\it l} - m)!}{({\it l} + m)!}
 \left(2-
 \delta_{0m}\right)\sum_{p=0}^{\it
 l}F^2_{{\it l}mp}(\inc)\sum_{q=-\infty}^{\infty}G^2_{{\it l}pq}(e)
 \;k_l(\omega_{lmpq})\;\cos\epsilon_l(\omega_{lmpq})
  ~_{\textstyle{_{\textstyle ~~,}}}~~~
 \label{5}
 \label{LLLL54}
 \label{L54}
  \ea
 where $\,l,\,m,\,p,\,q\,$ are integers, $\,F_{lmp}(\inc)\,$ are the inclination functions (Gooding \& Wagner 2008), $\,G_{lpq}(e)\,$ are
 the eccentricity polynomials coinciding with the Hansen coefficients $\,X^{\textstyle{^{(-l-1),\,(l-2p)}}}_{\textstyle{_{(l-2p+q)}}}(e)
 \,$, while the superscript ``{\it{sec}}" means: secular.

 The dynamical Love numbers $\,k_l\,$ and the phase lags $\,\epsilon_l\,$ are functions of the Fourier modes
 \ba
 \omega_{lmpq}\;\equiv\;({\it l}-2p)\;\dot{\omega}\,+\,({\it l}-2p+q)\;\dot{\cal{M}}\,+\,m\;(\dot{\Omega}\,-\,\dot{\theta})\,~,~~~
 \label{9}
 \ea
 $\theta\,$ being the primary's sidereal angle, and $\,\dot{\theta}\,$ being its angular velocity. While the tidal modes (\ref{9}) can be
 of either sign, the physical forcing frequencies
    \ba
    \chi_{lmpq}\,\equiv\,|\,\omega_{lmpq}\,|\,=\,|\,({\it l}-2p)\,\dot{\omega}\,+\,({\it l}-2p+q)\,\dot{\cal{M}}
    \,+\,m\,(\dot{\Omega}\,-\,\dot{\theta})~|
    \label{10}
    \ea
 at which the stress and strain oscillate in the primary are positive definite.

 A partial sum of series (\ref{LLLL54}), with $\,|{\it{l}}|,\,|q|,\,|j|\,\leq\,2\,$, was offered earlier by Darwin (1879). An explanation of Darwin's
 method in modern notation can be found in Ferraz-Mello, Rodr{\'{\i}}guez \& Hussmann (2008).\footnote{~In Ferraz-Mello, Rodr{\'{\i}}guez \& Hussmann
 (2008), the meaning of notations $\,\erbold\,$ and $\,\erbold^{\,*}\,$ is opposite to ours.}

 The power of the Darwin-Kaula approach lies in its compatibility with any rheology, i.e., with an arbitrary form of the mode-dependence of the
 products $\,k_l~\cos\epsilon_l\,$. It can be demonstrated that, for a homogeneous near-spherical primary, the functional form of the
 mode dependence of the product $\,k_l\,\cos\epsilon_{lmpq}\,$ is defined by index $\,l\,$ solely: $\,k_l(\omega_{lmpq})\,\cos\epsilon_{l}
 (\omega_{lmpq})\,$, the other three indices being attributed to the tidal mode. One can assume that this product depends not on the tidal
 mode $\,\omega_{lmpq}\,$ but on the positive definite physical frequency $\,\chi_{lmpq}\,$. This however will require some care in
 derivation of the tidal torque from the above expressions for the potential -- see Efroimsky (2012a,b) for details.

 We had to write down the secular part of the Kaula expansion of tide, because we shall use it as a benchmark wherewith to
 compare the empirical expression by MacDonald (1964), which we shall derive below in an accurate manner.

 Lacking the ability to accommodate an arbitrary rheology, the MacDonald approach (or the {\it{MacDonald torque}}) produces, after a
 necessary correction, a simple method which can sometimes be employed for getting qualitative understanding of the picture.

 \section{Tidal torque}

 Consider a secondary body of mass $\,M_{sec}\,$, located relative to its primary at $\,\erbold\,=\,(r,\,\lambda,\,\phi)\,$, where $\,
 \phi\,$ is the latitude, and $\,\lambda\,$ denotes the longitude reckoned from a meridian fixed on the primary. Let $\,U\,$ stand for
 the tidal-response amendment to the primary's potential. This amendment can be generated either by this secondary itself or by some
 other secondary of mass $\,M_{sec}^*\,$ located at $\,\erbold^{\,*}\,=\,(r^*,\,\lambda^*,\,\phi^*)\,$. In either case, the primary's
 tidal response to the gravity of the secondary will render a tidal force and torque acting on the secondary of mass $\,M_{sec}\,$.
 The torque's component perpendicular to the equator of the primary will be given by:
 \ba
 T_z\;=\;-\;M_{sec}\;\frac{\partial U(\erbold)}{\partial \lambda}\;\;\;.
 \label{T}
 \ea
 We would reiterate that (\ref{T}) is a component of the torque wherewith {\it{the primary acts on the secondary}} of mass $\,M_{sec}\,$.

 Then its negative will be the appropriate (i.e., orthogonal to the primary's equator) component of the torque wherewith {\it{the
 secondary acts on the primary:}}
 \ba
 {\cal{T}}_z(\erbold)\;=\;-\;T_z\;=\;\,M_{sec}\;\frac{\partial U(\erbold)}{\partial\lambda}~~~.~~~
 \label{15}
 \ea
 Derivation of formulae (\ref{T} - \ref{15}) is presented in Appendix \ref{tor}. These expressions are convenient when the
 tidal-response potential amendment $\,U\,$ is expressed through the spherical coordinates $~r\,,\,\lambda\,,\,\phi~$ and $~r^{\,*}\,,\,\lambda^{\,*}\,,\,\phi^{\,*}~$, as in formula (\ref{4}).

 Whenever the tidal response is expressed as a function of the orbital elements of the secondary and the sidereal angle $\,\theta\,$
 of the primary, it is practical to write down that the perpendicular-to-equator component of the torque acting on the primary as
 \ba
 {\cal{T}}_z(\erbold)\;=\;-\;M_{sec}\;\frac{\partial U(\erbold)}{\partial \theta}\;\;\;,
 \label{hh}
 \ea
 $\theta\,$ standing for the primary's sidereal angle (Efroimsky 2012a,b).

 We prefer to employ the terms {\emph{primary}} and {\emph{secondary}} rather than {\emph{planet}} and {\emph{satellite}}. This
 choice of terms is dictated by our intention to apply the below-developed machinery to research of tidal dissipation and spin
 evolution of a satellite. In this setting, the satellite is effectively playing the role of a tidally-distorted primary, its host
 planet acting as a tide-generating secondary. Whenever the method is applied to exploring the problem of planet despinning, the
 planet is understood as a primary, the host star being a tide-raising secondary.

 \section{
 MacDonald (1964)}\label{ger}

 When it comes to taking dissipation into account, expression (\ref{4}) turns out to be far more restrictive than (\ref{LLLL54}), in that
 (\ref{4}) becomes applicable to a very specific rheological model. This happens because a straightforward\footnote{~A more accurate
 treatment, which cannot be employed within the MacDonald model but is implementable within the Darwin-Kaula approach, is to expand
 the tide-raising potential $\,W\,$ and the tidal-response potential change $\,U\,$ into Fourier modes $\,\omega_{lmpq}\,$, and then to
 introduce the Love numbers $\,k_{lmpq}\,=\,k_l(\omega_{lmpq})\,$, phase lags $\,\epsilon_{lmpq}\,=\,\epsilon_l(\omega_{lmpq})\,$, and
 time lags $\,\Delta t_{lmpq}\,=\,\Delta t_l(\omega_{lmpq})\,$. This formalism (explained in detail in Efroimsky 2012a,b) enables one
 to express the so-introduced Love numbers through the rheological properties of the primary's mantle, and thereby to model adequately
 the frequency-dependence of these Love numbers.

 An intermediate, purely empirical, option would be to introduce ``Love numbers" $\,k_{lm}\,$, as if they were functions of both the degree
 $\,l\,$ and order $\,m\,$. This idea is implemented in the IERS Conventions on the Earth rotation (Petit \& Luzum 2011).
 In the LLR (Lunar Laser Ranging) integration software, tides in the Earth are parameterised by $\,k_{lm}\,$ and $\,\Delta t_{lm}\,$
 with $\,l\,=\,2\,$ and $\,m\,=\,0\,,\,1\,,\,2~~$ (Standish and Williams 2012).} option of instilling the
 delay into (\ref{4}) is to replace in this expression the perturber's coordinates $\,r^*(t)\,,\;\phi^*(t)\,,\;\lambda^*(t)\,$ with
 their delayed values, $\,r^*(t-\Delta t)\,,~\phi^*(t-\Delta t)\,,~\lambda^*(t-\Delta t)\,$. For example, instead of $\cos m (\lambda-
 \lambda^*)$ we should employ
 \ba
 \cos\left(~m~\lambda~-~m~\lambda^{\textstyle^{*^{\,\textstyle{^{(delayed)}}}}}~\right)
 ~=~\cos \left(\;m\;\lambda\;-\;(\,m\lambda^{\textstyle^*
 }\,-\,m\stackrel{\centerdot}{\lambda}{^{\textstyle^*}}\Delta t~)~\right)
 \;\;\;,
 \label{trick}
 \ea
 insertion whereof into (\ref{4}) yields:
  \ba
  \nonumber
 U(\erbold)=\,-\,\sum_{{\it{l}}=2}^{\infty}
   \frac{\;k_{\it l}\;{G\,M^*_{sec}}\;\,R^{\textstyle{^{2\it{l}+1}}}}{
  ~r(t)^{\textstyle{^{\it{l}+1}}}\;\;\;{r^{\;*}}(t-\Delta t)^{\textstyle{^{\it{l}+1}}}\;}
  \sum_{m=0}^{\it l}\frac{({\it l} - m)!}{({\it l} + m)!}
 \left(2  \right. ~~~~~~~~~~~~~~~~~~~~~~~~~~~~~~~~~~~
 \ea
 \ba
  \left.~~~~~~~~~~~~ -\delta_{0m}\right)P_{{\it{l}}m}(\sin\phi(t))P_{{
 \it{l}}m}(\sin\phi^*(t-\Delta t))\;\cos m(\lambda-\lambda^*+
 \dot{\lambda}^*
 \,\Delta t)~.~~
 \label{rr}
 \ea
 Expressing the longitude via the true anomaly $\,\nu\,$,
 \ba
 \lambda\;=\;-\;\theta\;+\;\Omega\;+\;\omega\;+\;\nu\;+\;O(\inc^2)\;=\;
 -\;\theta\;+\;\Omega\;+\;\omega\;+\;{\cal{M}}\;+\;2\;e\;\sin{\cal{M}}\;+
 \;O(e^2)\;+\;O(\inc^2)\;\;\;,\;\;\;\;
 \label{38}
 \ea
 and neglecting the apsidal and nodal precessions, we obtain:
 \ba
 \cos \left(\;(\,m\;\lambda\;-\;m\lambda^{\textstyle^*}\,)\;+\,m
 \dot{\lambda}^*
 \,\Delta t\;\right)\;=\;
 \cos \left(\;(\,m\;\lambda\;-\;m\lambda^{\textstyle^*}\,)\;+\,
 m\;(\dot{\nu}^*\,-\;\dot{\theta}^*)\;\Delta t\;+O~({i^*}^2)\;\right)\quad\quad
 \label{39}
 \ea
 or, in terms of the mean anomaly:
 \ba
 \nonumber
 \cos \left((m\lambda-m\lambda^{\textstyle^*})+m\stackrel{\centerdot}{\lambda}{^{*}}\,\Delta t\right)=
 \quad~\quad~\quad~\quad~\quad~\quad~\quad~\quad~\quad~\quad~\quad~\quad~\quad~\quad~\quad~\quad~\quad~\quad~\quad~\quad~\quad
 ~\\  \nonumber\\
 \cos \left((m\lambda-m\lambda^{\textstyle^*})+m(n^*-\dot{\theta}^*)\Delta t+2me^*n^*\Delta t\,\cos{\cal M}^*
 +~O({e^*}^2)~+O~({i^*}^2)~+~O(i^2)\,\right)~~~.~\quad
 \label{40}
 \label{4446}
 \ea
 This enables us to write down the potential as
 \ba
 \nonumber
 U(\erbold)=\,-\,G\,M_{sec}^*\,\sum_{{\it{l}}=2}^{\infty}k_l~
 \frac{R^{
 \textstyle{^{2\it{l}+1}}}}{r(t)^{
 \textstyle{^{\it{l}+1}}}{r^{^*}}(t-\Delta t)^{
 \textstyle{^{\it{l}+1}}}}\,
 \sum_{m=0}^{\it l}\frac{({\it l} - m)!
 }{({\it l} + m)!}\; \left(\,2\;-\;\delta_{0m}\right) P_{{\it{l}}m}(\sin\phi(t)\,)~~~~~~~~~~~~~~~~~~~~~~~~~~~~~~~~~~~~~~~~~~~~~~~~~~~~~~~~
 \ea
 \ba
 P_{{\it{l}}m}\left(\sin\phi^*\left(t-\Delta t\right)\,\right)\;\cos\left(\,m\,(\lambda-
 \lambda^{\textstyle^*})+\,m\,(\dot{\nu}^*-\dot{\theta}^*)\,\Delta t
 \,+O(\inc^2)+O({\inc^*}^2)
 \,\right)~\,,~~
 \label{837}
 \ea
 where $~P_{{\it{l}}m}(\sin\phi(t)\,)P_{{\it{l}}m}(\sin\phi^*(t-\Delta t)\,)~$ may be replaced, in the order of $~O(\inc^2)+O({\inc^*
 }^{2})+O(\inc\inc^*)\;$, with $\;P_{{\it{l}}m}(0)P_{{\it{l}}m}(0)\;\,$.
 A further simplification can be achieved by taking into account the $\;{\it l}\,=\,2\;$ contribution only. Omitting the $\,\lambda$-independent term with $\,m=0\,$ (in the expression for the torque, $\,m\,$
 will become a multiplier after the differentiation of $\,U(\erbold)\,$ with respect to $\,\lambda\,$), and omitting the $\,m=1\,$ term
 (as $\;P_{21}(0)\,=\,0\;$), we arrive at the expression
 \ba
 \nonumber
 U(\erbold)=\;-\;\frac{3}{4}\;
 \frac{G\;M_{sec}^*\;k_{2}\;R^{\textstyle{^5}}}{r(t)^{\textstyle{^{3}}}{
 r^{^*}}(t-\Delta t)^{\textstyle{^{3}}}}\;\left[1\;+\;O(\inc^2)+O({\inc^*}^2)+O(\inc\inc^*)
 \right]\;\cos \left(\;(\,2\;\lambda\;-\;2\lambda^{\textstyle^*}\,)
 \right.
 \ea
 \ba
 \left. ~~~~~~~~~~~~~~~~~~~~~~~~~~~~~~~~ +\;2\;(\dot{\nu}^*\,-\;\dot{\theta}^*)\;\Delta
 t~+~O(\inc^2)~+~O({\inc^*}^2)
 \;\right)~~~.~~~~
 \label{MacDonald}
 \ea
 In the special case when the tide-raising satellite is the same body as the tidally-perturbed one (i.e., when $\,r(t)=r^*(t)
 \,$, $\,M_{sec}=M_{sec}^*\,$, and $\,\lambda=\lambda^*\,$), this expression happens to coincide, in the leading order of $\,\inc\,$,
 and $\,\inc^*\,$,
 with the potential employed by MacDonald (1964). Thus we have reproduced his empirical approach, by starting with the rigorous formula
 (\ref{4}), and by performing the following sequence of approximations:

 \begin{itemize}

 \item{} First, when accommodating dissipation, we set all the time delays to be equal, for all the physical frequencies
 $\,\chi_{\textstyle{_{\it{l}mpq}}}\equiv\,|\,\omega_{\textstyle{_{\it{l}mpq}}}\,|\,$ involved in the tide:
 \ba
 \Delta t_{\textstyle{_{\it{l}mpq}}}\;\equiv\;\Delta t(\chi_{\textstyle{_{\it{l}mpq}}})\;=\;\Delta t\;\;\;.
 \label{set}
 \ea
 This point is explained in great detail in Efroimsky \& Makarov (2013).

 \item{} Second, we assume the smallness of the inclinations and lag, through the neglect of the relative errors
 $\,O(\inc^2)~$, $~O({\inc^*}^2)~$, and $~O(\inc\inc^*)$
 .\\

 \item{} Third, we truncate the series by leaving only the $\,{\it{l}}\,=\,m\,=\,2\,$ term.\\

 \end{itemize}

\noindent
 These three steps take us from (\ref{4}) to the
 approximation
 \begin{subequations}
 \ba
 U(\erbold)\approx\;-\;\frac{3}{4}\;
 \frac{G\;M_{sec}^*\;k_{2}\;R^{\textstyle{^5}}}{r(t)^{\textstyle{^{3}}}{
 r^{^*}}(t-\Delta t)^{\textstyle{^{3}}}}\;\cos \left(\,2\;\lambda\;-\;2\lambda^{\textstyle^*}~+~\epsilon~\right)\,~,
 \label{MacDonald_1_a}
 \ea
 With the tide-raising secondary set to coincide with the one perturbed by the tides on the primary (so $\,r(t)=r^*(t)\,$, $\,M_{sec}=M_{sec}^*\,$, and $\,\lambda=\lambda^*\,$), the above expression assumes the form of
 \ba
 U(\erbold)\approx\;-\;\frac{3}{4}\;
 G\;M_{sec}\;k_{2}\;\frac{\,R^{\textstyle{^5}}\,}{\,r(t)^3~r(t-\Delta t)\,}\;\,\cos \epsilon\,~.
 \label{giog}
 \ea
 In the denominator,  $\,r(t-\Delta t)\,$ can be replaced,
 \footnote{~\label{fufu}From
 $\,r=a(1-e^2)/(1+e\,\cos\nu)\,$ and $\,\partial\nu/\partial M=(1+e\,\cos\nu)^2/(1-e^2)^{3/2}\,$ it is straightforward that
 \ba
 \nonumber
 \Delta r\equiv r(t)-r(t-\Delta t)=-\frac{a\,e\,(1\,-\,e^2)}{(1\,+\,e~\cos\nu)^2}~\sin\nu~\Delta\nu\,+\,O\left(e\,(\Delta\nu)^2\,\right)
 \,=\,-\,\frac{a\,e\;\sin \nu}{(1-e^2)^{1/2}}\,n\,\Delta t\,+\,O\left(e\,(n\;\Delta t)^2\,\right)~~~.
 \ea
 Thus our replacement of $\,r^*(t-\Delta t)\,$ with $\,r^*(t)\,=\,r(t)\,$ entails a relative error of order $\,O(en\Delta t)\,$.
 In expressions (\ref{rr}), (\ref{837}), (\ref{MacDonald}), and (\ref{MacDonald_1}), the absolute error will be of the same order,
 for $\,\lambda\neq\lambda^*\,$. However for $\,\lambda=\lambda^*\,$ the absolute error will become $\,O(enQ^{-1}\Delta t)\,$, since
 $\,\sin\epsilon\,$ is of the same order as the inverse quality factor $\,Q\,$. In our estimates of errors, it is irrelevant whether we
 define $Q$ as that appropriate to the principal tidal mode or via formula (\ref{abandon}) below; so we simply use the generic
 notation $Q$.

 As we explained in Efroimsky \& Williams (2009), after averaging over one
 revolution of a nonresonant secondary about the primary, the absolute error reduces to $\,O\left(e^2n^2Q^{-2}(\Delta t)^2\right)~$.
 In a resonant case, though, we cannot enjoy this reduction, because in this case the averaging procedure looks different as the torque
 changes its sign over a period of the moon's revolution. For this reason, the absolute error remains $\,O(enQ^{-1}\Delta t)\,$.}
 in the order of $\,O(e^*n^*\Delta t)\,$, with $\,r^*(t)\,$.  With this simplification implemented, we end up with
 \ba
 U(\erbold)\approx\;-\;\frac{3}{4}\;
 G\;M_{sec}\;k_{2}\;\frac{\,R^{\textstyle{^5}}\,}{\,r^6\,}\;\,\cos \epsilon\,~.
 \label{MacDonald_1_b}
 \ea
 \label{MacDonald_1}
 \end{subequations}
 Expressions (\ref{MacDonald}) and (\ref{MacDonald_1}) contain the {\it{longitudinal lag}}
 \begin{subequations}
 \ba
 \epsilon\;\equiv\;m\,\stackrel{\centerdot}{\lambda}{^{\textstyle^*}}\Delta t\;=\;
 2\,(\dot{\nu}^*\,-\,\dot{\theta})\,\Delta t\;+\;O(\inc^2)\,~,
 \label{181818_a}
 \label{18_a}
 \ea
 $\,\nu^*\,$ and $\,\theta\,$ being the true anomaly of the perturber and the sidereal angle of the primary. In the special case
 (\ref{giog} - \ref{MacDonald_1_b}), when the tide-generating secondary and the secondary disturbed by the tide on the primary are one and the same
 body, the asterisks may be dropped:
 \ba
 \epsilon\;\equiv\;m\,\stackrel{\centerdot}{\lambda}\Delta t\;=\;
 2\,(\dot{\nu}\,-\,\dot{\theta})\,\Delta t\;+\;O(\inc^2)\,~.
 \label{181818_b}
 \label{18_b}
 \ea
 \label{181818}
 \label{18}
 \end{subequations}
 The spin rate $\,\dot{\theta}\,$ is a slow variable,
 in that it may be assumed constant over one orbiting cycle. The true anomaly is a fast variable. For a nonvanishing
 eccentricity, $\,\dot{\nu}\,$ too is a fast variable, and so is the lag $\,\epsilon\,$. This means that we should take into
 account these two quantities' variations over an orbital period.

 Expression (\ref{giog}) coincides, up to an irrelevant constant,\footnote{~The additional tidal potential $\,U\,$, as given by equation (21) in the work by MacDonald (1964), fails to vanish
 in the limit of zero geometric lag $\,\delta\,$. This minor irregularity, though, does not influence MacDonald's calculation of the tidal torque.} with the leading term of the appropriate formula from MacDonald (1964, eqn 21).
 To appreciate this fact, notice that, up to $\,O(\inc^2)\,$, the absolute value of the longitudinal lag $\,\epsilon\,$ is the double of the geometrical angle $\,\delta =|(\dot{\theta}-\dot{
 \nu})\,\Delta t|\,$ subtended at the primary's centre between the directions to the secondary and to the bulge,\footnote{~Be mindful
 that the double of the geometrical angle is {\it{not}} equal to the absolute value of $\,\epsilon_{\textstyle{_{2200}}}\,=\,2(n\,-\,
 \dot{\theta})\,\Delta t_{\textstyle{_{2200}}}\,$.} provided $\,\Delta t\,$ is postulated to be the same for all tidal modes.

 The analogy between the MacDonald theory and that of Darwin and Kaula can be traced also by starting from the series (\ref{L54}).
 Consider the case of the tidally perturbed secondary coinciding with the tide-raising one, so $\,\lambda\,=\,\lambda^*\,$, and all the orbital variables are identical to their
 counterparts with an asterisk. It will then be easy to notice that, formally (just formally), expression (\ref{MacDonald}) mimics the
 principal term of the series (\ref{LLLL54}), provided in this term the multiplier $\,G^2_{200}\,$ is replaced with unity, and the principal
 phase lag $\,\epsilon_{\textstyle{_{\textstyle{_{2200}}}}}\,\equiv\,2\,(n-\dot{\theta})\,\Delta t_{\textstyle{_{\textstyle{_{2200}}}}}\,$
 is replaced with the longitudinal lag (\ref{18}). This way, within the MacDonald formalism, the longitudinal lag (\ref{18}) is playing
 the role of an {\emph{instantaneous}} phase lag associated with double the {\emph{instantaneous}} synodic frequency
 \ba
 \chi\;=\;2\;|\,\dot{\nu}\;-\;\dot{\theta}\,|\;\;\;,
 \label{duga}
 \ea
 which is, up to $\,O(\inc^2)\,$, the double of the angular velocity wherewith the point located under the secondary (with the same
 latitude and longitude) is moving over the surface of the primary.

 To extend further the analogy between the MacDonald and Darwin-Kaula models, one can {\emph{define}} an auxiliary quantity
 \ba
 ``Q"\,=\,\frac{1}{\,\sin |\epsilon |\,}
 \label{abandon}
 \ea
 and derive from (\ref{181818}) and (\ref{abandon}) that, in the leading order of $\,\epsilon\,$ and
 $\,\inc\,$, this quantity satisfies
 \begin{subequations}
 \ba
 ``Q"\,=\,\frac{1}{\chi\;\Delta t}\;\;\;.
 \label{model}
 \ea
 A popular fallacy would then be to interpret (\ref{model}) as a rheological scaling law $\,Q\,\sim\,\chi^{\alpha}\,$ with $\,\alpha=-1\,$. That this
 interpretation is generally incorrect follows from the fact that the quantity $\,``Q"\,$, {\emph{defined}} through (\ref{abandon}), is
 {\underline{not}} obliged to coincide with the quality factor.\footnote{~\label{fufufu}Within the Darwin-Kaula theory, for each tidal mode
 $\,\omega_{{\it l}mpq}\,$, we introduce the phase lag as $~\epsilon_{{\it l}mpq}\,\equiv\,\omega_{lmpq}\,\Delta t_{{\it l}mpq}~$. Then we
 introduce the appropriate quality factor $\;Q_{{\it l}mpq}\,\equiv\,Q(\chi_{{\it l}mpq})\,=\,Q(\,|\omega_{{\it l}mpq}|\,)\;$ via the
 expression for the one-cycle energy loss:
 \ba
 \nonumber
 \Delta E_{cycle}(\chi_{\textstyle{_{lmpq}}})\;=\;-\;\frac{2\;\pi\;E_{peak}(\chi_{\textstyle{_{lmpq}}})}{Q
 (\chi_{\textstyle{_{lmpq}}})}\;\;\;.
 \label{defdef}
 \ea
 Finally, using the fact that $\,\chi_{\textstyle{_{lmpq}}}\equiv|\omega_{\textstyle{_{lmpq}}}|\,$ is the frequency of a
 {\emph{sinusoidal}} load, we prove that the afore introduced $\,\epsilon_{\textstyle{_{lmpq}}}\,$ and
 $\,Q_{\textstyle{_{lmpq}}}\,$ are interconnected as $\,1/Q=\sin\epsilon\,$, if $\,E_{peak}\,$ denotes the maximal energy, or in a
 more complex way, if $\,E_{peak}\,$ stands for the maximal work (Efroimsky 2012a,b).

 Within the MacDonald method, original or corrected as (\ref{set}), it is not {\emph{a priori}} clear if the overall peak work (or the
 overall peak energy stored) and the overall energy loss over a cycle are interconnected via the auxiliary quantity $\,``Q"\,$ in exactly
 the same manner as $\,E_{peak}(\chi_{\textstyle{_{lmpq}}})\,$ and $\,\Delta E_{cycle}(\chi_{\textstyle{_{lmpq}}})\,$ are interconnected
 by $\,Q(\chi_{\textstyle{_{lmpq}}})\,$ in the above expression. (Recall that the total cycle of the tidal
 load is, generally, nonsinusoidal.) Whenever we can prove that
 \ba
 \nonumber
 \Delta E_{cycle}^{(overall)}\;=\;-\;\frac{2\;\pi\;E_{peak}^{(overall)}}{``Q"}\;\;\;,
 \label{instant}
 \ea
 our $\,``Q"\,$ can be spared of the quotation marks and can be called {\emph{instantaneous}} quality
 factor, while (\ref{model}) can be treated as a reasonable approximation to scaling law
 $\,Q\sim\chi^{\alpha}\,$ with $\,\alpha=-1\,$. However, this should be justified in each particular
 case, not taken for granted.\label{ig}}

 To sidestep these difficulties, it would be safer to write the constant-time-lag rheological law, for small lags, simply as
 \ba
 |\,\epsilon\,|\;=\;\chi\;\Delta t~~~.
 \label{model_2}
 \ea
 \label{mod}
 \end{subequations}
 Historically, MacDonald (1964) arrived at his model via empirical reasoning. He certainly realised that the model was applicable to
 low inclinations  only. At
 the same time, this author failed to notice that the model also implied the frequency-independence of the time lag. In fact,
 this frequency-independence, (\ref{set}), is necessary to derive the MacDonald model (\ref{MacDonald_1}) from
 the generic expression (\ref{4}) for the tidal amendment to the primary's potential. This way, equality (\ref{set}), is {\it{a
 priori}} instilled into the model. In other words, the MacDonald model of tides includes in itself the rheological
 scaling law (\ref{model_2}) with a constant $\,\Delta t\,$. Unaware of this circumstance, MacDonald (1964)
 set the angular lag to be a frequency-independent constant, an assertion equivalent to the time lag scaling as inverse tidal frequency.
 However, as we just saw above, a consistent derivation of the MacDonald tidal model requires that the time lag be set
 frequency-independent.\footnote{~In application to a non-resonant setting, a constant-$\Delta t$ approach was taken yet by Darwin
 (1879). Later, MacDonald (1964) abandoned this method in favour of a constant-geometric lag calculation.
 Soon afterwards, though, Singer (1968) advocated for reinstallment of the constant-$\Delta t$ method. Doing so, he was motivated by an
 apparent paradox in MacDonald's treatment. As explained by Efroimsky \& Williams (2009), the paradox is nonexistent. Nonetheless, the
 work by Singer (1968) was fruitful at the time, as it renewed the interest in the constant-$\Delta t$ treatment. The full might of the
 model was revealed by Mignard (1979, 1980, 1981) who used it to develop closed expressions for the tidal force and torque.}
 Thus, to be consistent, the MacDonald method must be corrected by applying (\ref{set}) or, equivalently, (\ref{model_2}). In greater
 detail, the necessity of this amendment is considered in Efroimsky \& Makarov (2013).

 Even after the model is combined with rheology (\ref{set}), predictions of such a theory are of limited use. The problem is that this
 rheology is radically different from the actual behaviour of solids. As a result, calculations relying on (\ref{set})
 render implausibly long times of tidal despinning -- see, for
 example the discussion and references in Castillo-Rogez et al. (2011). This tells us that in realistic settings the MacDonald-style
 approach (\ref{MacDonald}) based on (\ref{set}) is inferior, compared to the Darwin and Kaula method (\ref{5}), which may, in principle,
 be applied to any rheology. Despite this, the MacDonald torque remains a convenient toy model, capable of furnishing results which
 are qualitatively acceptable over not too long timescales (Hut 1981, Dobrovolskis 2007).


  \section{The MacDonald torque and the ensuing dynamical model by Goldreich.
  The case of an oblate body}\label{ek}

 In this section, we shall trace how the MacDonald theory of bodily tides yields the model of near-resonance spin dynamics by
 Goldreich (1966). Then we shall recall an oversight in the MacDonald theory, and shall demonstrate that correction of that
 oversight brings a minor alteration into Goldreich's model of spin evolution.

 The goal of this section is limited to consideration of the tidal torque solely. So we assume that the body is oblate, and the
 triaxiality-caused torque does not show up. Appropriate for spinning gaseous objects, this treatment indicates that the tidal torque
 stays finite for synchronous rotation and vanishes at an angular velocity slightly faster than synchronous.

 In Section \ref{JW} below we shall address the more general setting appropriate to telluric bodies, with triaxiality included. While
 some authors (e.g., Heller et al. 2011) ignore the triaxiality-caused torque in their treatment of solid planets, it turns out that inclusion of the permanent-figure torque renders important physical consequences and changes the picture completely,
 as will be seen in Section \ref{JW}.

 \subsection{The MacDonald torque}\label{restrict}

 We shall restrict ourselves to the case of the tidally perturbed secondary coinciding with the tide-raising one, so
 $\,{{M}}_{sec}\,=\,{{M}}_{sec}^*\,$, and all the orbital variables are identical to their counterparts with an asterisk.

 Differentiating (\ref{MacDonald}) with respect to $\,\lambda\,$, and then setting $\,\lambda\,=\,\lambda^*\,$, we obtain the following
 expression for the polar component of the torque, for low $\,\inc\;$:
 \ba
 \nonumber
 {\cal{T}}_z &=& \frac{3}{2}~{G\,M_{sec}^2}\;k_{2}~
 \frac{R^{\textstyle{^5}}}{~r^{\textstyle{^{3}}}(t)\;\;\;{
 r^{\;*}}^{\textstyle{^{3}}}(t-\Delta t)\;} \,\;\sin
 \left(\,2\,(\dot{\nu}\,-\,\dot{\theta})\,\Delta t\,\right)
 +O(\inc^2/Q)\\
 \label{17}\\
 \nonumber
 &=&\frac{3}{2}~{G\,M_{sec}^2}\;k_{2}~
 \frac{R^{\textstyle{^5}}}{r^{\textstyle{^6}}}~\,\sin
 \left(\,2\,(\dot{\nu}\,-\,\dot{\theta})\,\Delta t\,\right)
 +O(\inc^2/Q)+O(en\Delta t/Q)\;\;,~~
  \ea
 where the error $\,O(en\Delta t/Q)\,$ emerges when we identify the lagging distance $\,r^*(t-\Delta t)\,$ with $\,r^*(t)\,=\,r(t)
 \,$, as explained in footnote \ref{fufu} in the preceding section. Referring the Reader to Efroimsky \& Williams (2009) for this and
 other technicalities, we would mention that (\ref{17}) is equivalent to the Darwin torque only under the condition that the
 rheological model (\ref{set} - \ref{model}) is accepted. For potentials, employment of model (\ref{set} - \ref{model}) enables one
 to wrap up the infinite series (\ref{5}) into the elegant finite form (\ref{rr}). For torques (truncated to $\,{\it{l}}=2\,$ only),
 similar wrapping of the appropriate series is available within the said model.

 In the preceding section, we explained the geometric meaning of the longitudinal lag (\ref{18}): its absolute value is the double of
 the geometric angle separating the directions to the bulge and the secondary as seen from the primary's centre.

 If we {\emph{define}} a quantity $\,``Q"\,$ via (\ref{model}), the MacDonald torque will look:
 \ba
 \nonumber
 {\cal{T}}_z&=&\frac{3}{2}~G~M_{sec}^{\,2}\;k_{2}\frac{R^{\textstyle{^5}}}{~r^{\textstyle{^{3}}}(t)\;\;\;{
 r^{\;*}}^{\textstyle{^{3}}}(t-\Delta t)\;}\;\,\sin\epsilon+O(\inc^2/Q) ~~~~~~~~~~~~~~~~~~~~~~~~~~~~~~~~~~~\\
  \label{19}\\
  \nonumber
 &=&\frac{3}{2}~{G~M_{sec}^{\,2}}\;k_{2}~\frac{R^{\textstyle{^5}}}{r^{\textstyle{^{6}}}
 } ~\frac{1}{``Q"}\,~\mbox{sgn}(\dot{\nu}-\dot{\theta})+O(\inc^2/Q) +O(en\Delta t/Q)
 ~~~.~~~~~~~~
 \ea
 For a nonzero eccentricity, the quantity $\,``Q"\,$ should not be interpreted as an instantaneous quality factor, because it is
 {\it{not}} guaranteed to interconnect the peak work or peak energy and the one-cycle energy loss in a manner appropriate to a
 quality factor -- see footnote \ref{fufufu} in the previous section. Therefore a more reasonable and practical way of writing the
 MacDonald torque (\ref{17}) would be through using (\ref{model_2}) or
 (\ref{181818}):
 \ba
 \label{1919}
 {\cal{T}}_z~=~\frac{3}{2}~{GM_{sec}^2}~k_{2} \frac{R^{\textstyle{^5}}}{r^{\textstyle{^{6}}}
 }~\Delta t~2~(\dot{\nu}-\dot{\theta})~+~O(\inc^2/Q)~+~O(en\Delta t/Q)
 ~~~.~~~~~~~~
 \ea
 As we saw above, the MacDonald model is self-consistent only for $\,\Delta t\,$ being a frequency-independent
 constant. However the factor $~\dot{\nu}-\dot{\theta}~$ showing up in (\ref{1919}) varies over a cycle, for which reason
 the torque needs averaging.
 This averaging is carried out in Appendix \ref{averaging}. In the vicinity of the $\,1:1\,$ resonance
 (for $\,\dot{\theta}\,$ close to $\,n\,$), the sign of $~\dot{\nu}-\dot{\theta}~$ changes twice over a cycle, which makes
 the averaging procedure nontrivial. This situation is to be addressed in subsections \ref{incorrect} and \ref{correct}.

  \subsection{Goldreich (1966): treatment based on the MacDonald torque}\label{incorrect}

 In this subsection, we shall briefly recall a presently conventional method pioneered almost half a century ago by Goldreich
 (1966). Based on the MacDonald tide theory, this method has inherited both its simplicity and its flaws.

 The 1:1 resonance takes place when the spin rate $\,\dot{\theta}\,$ of the primary (the satellite) is equal to the mean motion
 $\,n\,$ wherewith the secondary body (the planet) is apparently orbiting the primary. The formula (\ref{19}) for the MacDonald
 torque contains not the difference $\,\dot{\theta}\,-\,n\,$ but the difference $\,\dot{\theta}\,-\,\dot{\nu}\,$, for which
 reason the expression under the integral may twice change its sign in the course of one revolution.

 With aid of the formula
 \ba
 \nu\;=\;{\cal{M}}\;+\;2\;e\;\sin {\cal{M}}\;+\;\frac{5}{4}\;e^2\;\sin 2{\cal{M}}\;+\;O(e^3)\;\;\;
 \label{eh}
 \ea
 and under the assumption that $\dot{e}\ll n$, we shape the difference of
 our concern into the form of
 \ba
 \nonumber
 \dot{\theta}\;-\;\dot{\nu}\;=\,\left(\dot{\theta}-n\right)\;-\;2\;n\;e\;\left(\cos {\cal{M}}
 \;+\;\frac{5}{4}\;e\;\cos 2{\cal{M}}\right)\;+\;O(e^3)
 \label{}
 \ea
 \ba
 \left.~~~~~~~~~~~~\right.=\;\dot{\eta}\;-\;2\;n\;e\;\left(\,\cos {\cal{M}}
 \;+\;\frac{5}{4}\;e\;\cos 2{\cal{M}}\,\right)\;+\;O(e^3)\;\;\;
 \label{difference}
 \ea
 or, equivalently,
 \ba
 \frac{\dot{\theta}\;-\;\dot{\nu}}{2\;n\;e}\;=\;-\;\left[\;\cos {\cal{M}}
 \;+\;\frac{5}{4}\;e\;\cos 2{\cal{M}}\;-\;\frac{\dot{\eta}}{2\;n\;e}\;+\;O(e^2)\,\right]\;\;\;.
 \label{difference2}
 \ea
 Here
 \ba
 {\eta}\;\equiv\;\theta\;-\;{\cal{M}}\;-\;\omega\;-\;\Omega\;\;\;
 \label{gede}
 \ea
 is a slowly changing quantity, whose time-derivative $\;\dot{\eta}\;\equiv\;\dot{\theta}\;-\;n\;$ becomes nil when the system goes
 through the resonance.\footnote{~Goldreich (1966) defined this quantity simply as
 ${\eta}\;\equiv\;\theta\;-\;{\cal{M}}\;,$
 because he reckoned $\,\theta\,$ from from a fixed perihelion direction. We however reference
 our $\,\theta\,$ from a direction fixed in space. (It is the same direction wherefrom the node
 is reckoned.) This convention originates from our
 definition of $\,\theta\,$ as the sidereal angle -- it is in this capacity that $\,\theta\,$
 was introduced back in equations (\ref{9} - \ref{10}). As we are
 not considering the nodal or apsidal precession, our subsequent formulae containing
 $\,\dot{\theta}\,$ will be equivalent to those ensuing from Goldreich's definition of
 $\,\theta\,$.} To impart the words ``slowly changing" with a definite meaning, we assert that $\;\dot{\eta}/n\;$ is of
 order $\,e^2\,$  -- a claim to be justified {\emph{a posteriori}}.

 Expression (\ref{difference}) changes its sign at the points where
 \ba
 \cos {\cal{M}}\;=\;-\;\frac{5}{4}\;e\;\cos
 2{\cal{M}}\;+\;\frac{\dot{\eta}}{2\;n\;e}\;+\;O(e^2)\;\;\;.
 \label{condition}
 \ea
 As all the terms on the right-hand side of (\ref{condition}) are of order $~e~$ at most, so must be the term on the left-hand side. Hence
 condition (\ref{condition}) is obeyed in the two points whose mean anomaly (and therefore also true anomaly) is close to $\,\pm\,\pi
 /2\,$:
 \ba
 \nu\;=\;\pm\;\left(\,\frac{\pi}{2}\;-\;\delta\,\right)\;\;\;,
 \label{fo}
 \ea
 $\delta\,$ being of order $\,e\,$. From (\ref{fo}) and (\ref{eh}) we obtain:
 \ba
 \sin\delta\;=\;\cos\nu\;=\;\cos\left(\,{\cal{M}}\;+\;2\;e\;\sin{\cal{M}}\;+\;O(e^2)\,\right)\;=\;
 \cos{\cal{M}}\;-\;2\;e\;\sin^2{\cal{M}}\;+\;O(e^2)\;\;\;,\;\;\;
 \label{}
 \ea
 which, in combination with (\ref{condition}), entails:
 \ba
 \sin\delta\,=\,-\,\frac{5}{4}\,e\;\cos
 2{\cal{M}}\,+\,\frac{\dot{\eta}}{2\,n\,e}\,-\,2\,e\;\sin^2{\cal{M}}\,+O(e^2)\,
 =\,\frac{\dot{\eta}}{2\,n\,e}\,-\,e\,\left(\,\frac{5}{4}\,-\,\frac{1}{2}\;\sin^2{\cal{M}}\,\right)\,+O(e^2)\;\;\;.\;\;\;
 \label{for}
 \ea
 Insertion of (\ref{fo}) into (\ref{eh}) also yields $\,\;\pm\,\sin {\cal{M}}\,=\,\cos\delta\,+\,O(e)\,$. Recalling that $\,\delta\,$
 is of order $\,e\,$, we obtain: $\,\;\sin^2 {\cal{M}}\,=\,1\,-\,\sin^2\delta\,+\,O(e)\,=\,1\,+\,O(e)\,$.
 This enables us to rewrite (\ref{for}) as
 \ba
 \delta\;=\;\frac{\dot{\eta}}{2\,n\,e}\;-\;\frac{3}{4}\;e\;+\;O(e^2)\;\;\;.
 \label{fort}
 \ea

 Before finding the rate $\,\dot{\eta}\,\equiv\,\dot{\theta}\,-\,n~$ at which the resonance is traversed, let us enquire if
 perhaps it could be simply put nil, the satellite being permanently kept in the resonance. The answer is negative, because
 for a vanishing $\,\dot{\theta}\,-\,n\,$ the difference $\,\dot{\theta}\,-\,\dot{\nu}\,$ emerging in (\ref{difference})
 becomes a varying quantity of an alternating sign, and so becomes the torque. On general grounds, one should not expect that
 the average torque becomes nil for a vanishing $\,\dot{\eta}\,$, though it may vanish for some finite value of
 $\,\dot{\eta}\,$.

 To find this value of $\,\dot{\eta}\,$, many authors (Goldreich 1966, eqn.$\,$15; Kaula 1968, eqn$\,$4.5.29; Murray \&
 Dermott 1999, eqn.$\,$5.11) simply integrated the MacDonald torque (\ref{19}), assuming the quality factor $\,Q\,$ constant,
 and thus keeping it outside the integral:
 \ba
 \nonumber
 \langle{\cal{T}}_z\rangle^{{^{(Goldreich)}}}\;&=&\;-~\frac{3\,G\,M_{sec}^{\textstyle{^{\,2}}}\,k_{2}\,R}{4\;\pi\;a^2\;Q}
 \;\,\frac{1}{\,\left(1\;-\;e^2\right)^{1/2}\,}\;\int_{0}^{2\pi}\;
 \frac{R^{\textstyle{^4}}}{r^4}\;\,{\mbox{sgn}(\dot{\theta}\,-\,\,\dot{\nu})}
 \,\;{d\nu}\;\\
 \nonumber\\
 \nonumber\\
 \nonumber
 &=&\;-~\frac{3\,G\,M_{sec}^{\textstyle{^{\,2}}}\,k_{2}\,R}{4\;\pi\;a^2\;Q}
 \;\,\frac{2}{\,\left(1\;-\;e^2\right)^{1/2}\,}\;\left[\;\int_{0}^{\pi/2-\delta}\;
 \frac{R^{\textstyle{^4}}}{r^4}\;\,
 \,\;{d\nu}\;-\;\int_{\pi/2-\delta\;}^{\pi}\;
 \frac{R^{\textstyle{^4}}}{r^4}\;\,
 \,\;{d\nu}\;\right]
 \ea
 \ba
 ~~~~~~~~=~
 \frac{3\,G\,M_{sec}^{\textstyle{^{\,2}}}\,k_{2}\,R^5}{2\;\pi\;a^6\;Q}\,
 \left[\int_{0}^{\pi/2-\delta}-\int_{\pi/2-\delta}^{\pi}\right]\,
 (1+4e\;\cos\nu)\,
 {d\nu}+O(e^2)
  ~\\
  \nonumber\\
  \nonumber\\
 =~\frac{3~G~M_{sec}^{\textstyle{^{\,2}}}~k_{2}~R^5}{\pi\;a^6\;Q}\,
 ~(4e\;\cos\delta~-~\delta)~
 +~O(e^2)
 ~~.~~~~~~~~~~~~~~~~~~~~~
 \label{geg}
 \ea
 In a trapping situation, the average torque vanishes. So (\ref{geg}) entails:
 \ba
 \delta\;=\;4\;e\;\cos\delta\;=\;4\;e\;+\;O(e^2)\;\;\;.
 \label{}
 \ea
 By combining the latter with (\ref{fort}), the afore quoted authors arrived at
 \ba
 \dot{\eta}_{\textstyle{_{stall}}}\;=\;\frac{19}{2}\;n\;e^2\;\;\;,
 \label{oldie}
 \ea
 an expression repeated later in papers and textbooks (e.g., eqn. 4.5.37 in Kaula 1968, or eqn. 5.14 in Murray \& Dermott 1999).
 This result however needs to be corrected, because the MacDonald approach is incompatible with the frequency-independence of
 $\,Q\,$ assumed in (\ref{geg}).

 Be mindful that (\ref{oldie}), as well as its corrected version (\ref{kik}) to be derived below, indicate that $\,\dot{\eta}\,$ is of
 order $\,e^2\,$. This justifies our assertion made in the paragraph after formula (\ref{gede}).

 \subsection{The corrected MacDonald model}\label{correct}

 To impart the MacDonald treatment consistently, one has to calculate the averaged
 torque, with the frequency-dependence of $\,Q\,$ taken into consideration. As we
 saw in subsection \ref{ger}, the MacDonald approach fixes this dependence in a
 manner that can, with some reservations, be approximated with
 \ba
 Q\;=\;\frac{1}{\chi\;\Delta t}
 \label{q}
 \ea
 or, in more general notations,
 \ba
 Q\;=\;{\cal E}^{\alpha}\;\chi^{\alpha}~~~,~~~\mbox{with}~~~\alpha\;=\;-\;1~~~,
 \label{qu}
 \ea
 where the double instantaneous synodic frequency is given by (\ref{duga}).

 The form (\ref{qu}) of the rheological law is more convenient, as it leaves us an
 opportunity to consider values of $\,\alpha\,$ different from $\;-1\,$. For any
 value of $\,\alpha\,$, the constant $\,{\cal{E}}\,$ is an integral rheological
 parameter, which has the dimension of time, and whose physical meaning is
 discussed in Efroimsky \& Lainey (2007). It can be demonstrated that in the
 special case of $\,\alpha=-1\,$ the parameter $\,{\cal{E}}\,$ coincides with
 the time lag $\,\Delta t\,$. For actual terrestrial bodies, $\,\alpha\,$ is
 different from $~-\,1\,$, and the integral rheological parameter $\,{\cal{E}}\,$
 is related to the time lag in a more complicated manner ({\it{Ibid.}}).

 As demonstrated in Appendix \ref{averaging}, insertion of (\ref{qu}) and (\ref{duga}) into (\ref{19}), for $\,\alpha=-1\,$,
 or equivalently, direct employment of (\ref{1919}), entails the following expression for the orbit-averaged torque acting on a
 secondary:
 \ba
 \langle\;{\cal{T}}_z\;\,\rangle\;=\;-\;{\cal Z}\;\left[
 \;\dot{\theta}\;\,{\cal A}(e)\;-\;n\;{\cal N}(e)\;\right]+O(\inc^2/Q)+O(Q^{-3})+O(en\Delta t/Q)\;\;\;,
 \label{Laskar}
 \label{47}
 \label{21}
 \ea
 where
 \ba
 {\cal A}(e)\;=\;\left(\,1\;+\;3\;e^2\;+\;\frac{3}{8}\;e^4\,
 \right)\;\left(\,1\,-\,e^2\,\right)^{-9/2}~=~1~+~\frac{15}{2}~e^2~+~\frac{105}{4}~e^4~+~O(e^6)~~~~~~~~~~~~~~~~~
 \label{48}
 \label{22}
 \ea
 and
 \ba
 {\cal N}(e)\;=\;\left(\,1\;+\;\frac{15}{2}\;e^2\;+\;\frac{45}{8}\;
 e^4\;+\;\frac{5}{16}\;e^6\,\right)\;\left(\,1\,-\,e^2\,\right)^{-6}~=~1~+~\frac{27}{2}~e^2~+~\frac{573}{8}~e^4~+~O(e^6)~~~,\quad
 \label{49}
 \label{23}
 \ea
 while the factor $\,{\cal{Z}}\,$ is given by \footnote{~Recall that, for $\,\alpha=-1\,$, the rheological parameter
 $\,{\cal{E}}\,$ is simply the time lag $\,\Delta t$.}
 \ba
 {\cal Z}\,=\,\frac{3\,G\,M_{sec}^{\textstyle{^{\,2}}}\;\,k_{2}\;{\cal E}}{R}\;
 \frac{R^{\textstyle{^6}}}{a^6}\,
 =\,\frac{3\,n^2\,M_{sec}^{\textstyle{^{\,2}}}\;\,k_{2}\;{\Delta t}}{
 (M_{prim}\,+\,M_{sec})}\;\frac{R^{\textstyle{^5}}}{a^3}
 \,=\,\frac{3\,n\,M_{sec}^{\textstyle{^{\,2}}}\;\,k_{2}}{Q\;(M_{prim}\,+\,M_{sec})}\;\frac{R^{\textstyle{^5}}}{a^3}\;\,
 \frac{n}{\chi}\;\;\;,~~~~
 \label{54}
 \ea
 $M_{prim}\,$ and $\,M_{sec}\,$ being the masses of the primary and the secondary.

 Be mindful that the right-hand side of (\ref{54}) contains a multiplier $~\frac{\textstyle n}{\textstyle\chi}\,=\,\frac{\textstyle n}{
 \textstyle 2\,|\stackrel{\bf\centerdot}{\theta\,}-\stackrel{\bf\centerdot}{\nu\,}|}\,~$ which is missing in the despinning formula
 employed by Correia \& Laskar (2004, 2009). This happened because in {\it{Ibid.}} the quality factor was introduced as $\,1/(n\,\Delta
 t)\,$ and not as $\,1/(\chi\,\Delta t)\,$ -- see the line after formula (9) in Correia \& Laskar (2009). In reality, the quality factor
 $\,Q\,$ must, of course, be a function of the forcing frequency $\,\chi\,$ (which happens to coincide with the mean motion $\,n\,$ in
 the 3:2 and 1:2 resonances but differs from $\,n\,$ outside these).

 The quality factor $\,Q\,$ being (within this model) inversely proportional to $\,\chi\,$, the presence of the $\,\frac{
 \textstyle n}{\textstyle\chi}\,$ factor in (\ref{54}) makes the overall factor $\,{\cal{Z}}\,$ a frequency-independent constant.

 Rewriting (\ref{Laskar}) as
 \ba
 \nonumber
 \langle\,{\cal{T}}_z\,\rangle &=&
 -\,{\cal Z}\left[
 \dot{\theta}\,\left(1+\,\frac{15}{2}\;e^2+\,\frac{105}{4}\;e^4\right)-n\left(1+\,\frac{27}{2}\;e^2+\,
 \frac{573}{8}\;e^4\right)\right]+O(e^6)+O(\inc^2/Q)+O(en\Delta t/Q)\\
 \nonumber\\
 &=&-\,{\cal Z}\left[
 \dot{\eta}\,\left(1+\,\frac{15}{2}\;e^2+\,\frac{105}{4}\;e^4\right)-6ne^2\left(1+\,\frac{121}{16}\;e^2\right)\right]+O(e^6)+O(\inc^2/Q)+O(en\Delta t/Q)
  \;\;,\quad\quad\quad
 \label{Laskar_av}
 \ea
 we see that it vanishes when the rate of change of $\,{\eta}\,=\,{\theta}\,-\,{\cal{M}}\,-\,\omega\,-\,\Omega\;$
 accepts the value
 \ba
 \dot{\eta}_{\textstyle{_{stall}}}\,=
 \;{6}\;n\;e^2\,
  +\;\frac{3}{8}~n~e^4
  \,+~O(e^6)~+~O(\inc^2/Q)~+~
 O(en\Delta t/Q)\;\;\;.\;
 \label{kik}
 \ea
 For $\,\etadot\,$ larger or smaller than $\,\dot{\eta}_{\textstyle{_{stall}}}\,$, the average torque (\ref{Laskar_av}) is nonzero
 and impels $\,\etadot\,$ to evolve towards the stall value (\ref{kik}).

 On the right-hand side of (\ref{kik}), the leading-order term contains a numerical factor of $\,6\,$, as different from the factor
 $\,19/2\,$ showing up in (\ref{oldie}). The necessity to change $\,19/2\,$ to $\,6\,$ in the $\,e^2\,$ term was
 pointed out by Rodr{\'{\i}}guez, Ferraz-Mello \& Hussmann (2008, eqn.$\,$2.4), who had arrived at this conclusion through some
 different considerations (which, too, were based on the frequency-dependence (\ref{qu})$\,$). The same result can be obtained
 within the Darwin-Kaula approach, provided the scaling law (\ref{qu}) is employed (Efroimsky 2012a,b).$\,$\footnote{~Implicitly,
 this result is present also in Correia et al. (2011, eqn 20), Laskar \& Correia (2003, eqn 9), and in Hut (1981). The
 earliest implicit occurrence of this result was in Goldreich \& Peale (1966, eqn 24).}

 Although correction of the oversight in the MacDonald torque renders a number different from
 the one furnished by Goldreich's development (6 instead of 19/2), qualitatively the principal
 conclusion by Goldreich (1966) remains unchanged: when an oblate body's spin is evolving
 toward the resonance, vanishing of the average tidal torque entails spin slightly faster than
 resonant, a so-called {\it{pseudosynchronous rotation}}. It however should be strongly
 emphasised that the possibility of pseudosynchronous rotation hinges upon the dissipation
 model employed. Makarov and Efroimsky (2013) have found that a more realistic tidal
 dissipation model than the corrected MacDonald torque makes pseudosynchronous rotation
 impossible.

 \section{Evolution of rotation near the 1:1 resonance.\\ The case of a triaxial body}\label{JW}

 In distinction from a gaseous or liquid body, a solid body would be expected to have a permanent figure in addition to the tidal
 distortion discussed so far. Goldreich (1966) demonstrated that this permanent figure plays an important role in determining if a
 primary, whose rotation is being slowed down or sped up by the tidal torque caused by a secondary, can be captured into the
 synchronous rotation state. This section follows Goldreich's derivation, but substitutes the MacDonald tidal torque with its
 corrected version, and also uses a more general mass expression.

 \subsection{Rotating primary subject to a triaxiality-caused torque.\\
 The equation of motion and the first integral.}

 Consider a triaxial primary body, which has its principal moments of inertia ordered as $\,A\,<\,B\,<\,C\,$, and which is rotating
 about the maximal-inertia axis associated with moment $\,C\,$. About the primary, a secondary body describes a near-equatorial orbit
 (so its inclination on the primary's equator, $\,i\,$, may be neglected). The secondary exerts on the primary two torques. One being
 tidal, the other is triaxiality-caused, i.e., generated by the existence of the permanent figure of the primary. Its component acting
 on the primary about the maximal-inertia axis is
   \ba
   {\cal{T}}_{\textstyle_{triax}}\;=\;\frac{3}{2}\;(B~-~A)\;\frac{G~M_{sec}}{r^3}\;\,\sin 2\lambda~~~,
   \label{650}
   \ea
 the longitude $\,\lambda\,$ being furnished by formula (\ref{38}). In that formula, the sidereal angle $\,\theta\,$ is reckoned
 from a reference direction in space to the principal axis associated with moment $\,A\,$.

 The acceleration of the sidereal angle then obeys
   \ba
   C \;\ddot{\theta} - \frac{3}{2}\;(B~-~A)\;\frac{G~M_{sec}}{r^3}\,~\sin 2\lambda ~=~0~~~.
   \label{651}
   \ea
 In neglect of the nodal and apsidal precession, as well as of $\,\stackrel{\bf\centerdot}{{\cal{M}}_{\textstyle{_0}}}\,$, definition
 (\ref{gede}) yields:
  \footnote{~The caveat about three neglected items implies that our $\,n\,$ is the {\emph{osculating}} mean motion
  $\,n(t)\,\equiv\,\sqrt{\mu/a(t)^3}\,$, and that we extend this definition to perturbed settings. The so-defined mean anomaly evolves
  in time as $\,{\cal{M}}\,=\,{\cal{M}}_0(t)+\int_{t_o}n(t)\,dt\,$, whence $\,\dot{\cal{M}} =\dot{\cal{M}}_0+n(t)\,$.

  The said caveat becomes redundant when $\,n\,$ is defined as the {\emph{apparent}} mean motion, i.e., either as the mean-anomaly rate
  $\,d{\cal M}/dt\,$ or as the mean-longitude rate $\,dL/dt\,=\,d\Omega/dt\,+\,d\omega/dt\,+\,d{\cal{M}}/dt\,$ (Williams et al. 2001).
  While the first-order perturbations of $\,a(t)\,$ and of the osculating mean motion $\,\sqrt{\mu/a(t)^3}\,$ include no secular terms,
  such terms are often contained in the epoch terms $\,\dot{\Omega}\,$, $\,\dot{\omega}\,$, and $\,\dot{\cal{M}}_0\,$. This produces
  the difference between the apparent mean motion defined as $\,dL/dt\,$ (or as $\,d{\cal M}/dt\,$) and the osculating mean motion $\,
  \sqrt{\mu/a(t)^3}\,$.}
 \ba
 \etadot\;=\;\thetadot\;-\;n~~~.
 \label{653}
 \ea
 Ignoring changes in the mean motion,\footnote{~Our neglect of evolution of the mean motion is acceptable, because orbital acceleration
 $\,\dot{n}\,$ is normally much smaller than the spin acceleration $\,\ddot{\theta}\,$. Indeed, for torques arising from the primary, $\,{\textstyle\dot{n}}/{\textstyle
 \thetadotdot}\,$ is of the same order as the ratio of $\,C\,$ to the orbital moment of inertia.} we write:
 \ba
  \thetadotdot = \etadotdot
 ~~~,
 \label{654}
 \ea
 so the first term in (\ref{651}) becomes simply $\,C\ddot{\eta} \,$.

 To process the second term in (\ref{651}), we would compare (\ref{38}) with (\ref{gede}):
 \ba
 \nonumber
 \lambda ~=~ -~\theta~+~\Omega~+~\omega~+~\nu~+~O(i^2) &=& (\,-\,\theta~+~\Omega~+~\omega~+~{\cal{M}})~+~(\nu\,-\,{\cal{M}})~+~O(i^2)\\
 \nonumber\\
 &=& -~\eta~+~(\nu\,-\,{\cal{M}})~+~O(i^2)
 \label{s}
 \ea
 In neglect of the inclination, the following approximation is acceptable in the vicinity of the 1:1 resonance:\footnote{~Pioneered
 by Goldreich \& Peale (1966), the approximation is explained in more detail by Murray \& Dermott (1999) whose treatment omits terms of
 the order of $\,e^3\,$ and higher (see formulae 5.59 - 5.60 in {\it{Ibid.}}).
 To justify the omission, recall that before averaging the expansion of $~r^{-3}~{\sin 2\,\lambda}~$ includes a series of terms with
 different arguments of the form $\,\sin(2\eta+q{\cal M})\,$, where $\,q\,$ is an integer. Since we are interested in dynamics in the
 vicinity of the 1:1 resonance, where $\,\eta\,$ is a slow variable, then only the $\,\sin(2\eta)\,$ term remains after the average.
 The averaged-out terms are of the order of $\,e\,$ and higher powers including the $\,e^3\,$ terms.
 }
 \ba
 \langle\,r^{-3}~{\sin 2\,\lambda}\,\rangle~=~-~G_{200}(e)\,a^{-3}\,{\sin2\eta}~~~,
 \label{z}
 \ea
 where $\,\langle...\rangle\,$ signifies {\it{orbital}} averaging, while the eccentricity function can
 be approximated with
 \ba
 G_{200}(e)\,=~1~-~\frac{5}{2}~e^2\,+~O(e^4)~~~.
 \label{g}
 \ea
 This way, omitting $\,O(i^2)\,$ in (\ref{s}) and substituting $\,\sin 2\lambda\,$ with its average in
 (\ref{651}), we transform the second term in (\ref{651}) to:
 \ba
 +~\frac{3}{2}~(B\,-\,A)~\frac{G\;M_{sec}}{a^3}~G_{200}(e)~\sin 2\eta~~~.
 \label{x}
 \ea
 While Goldreich (1966) assumed that the mass of the secondary is much larger than the mass of the primary, we do not impose
 this restriction. Combining Kepler's third law, $~G(M_{sec}+M_{prim})/a^3=n^2~$, with formulae (\ref{651}), (\ref{654}), and
 (\ref{x}), we finally arrive at
   \ba
   C \etadotdot ~+~\frac{3}{2}~(B\,-\,A)~\frac{M_{sec}}{M_{sec}\,+\,M_{prim}}~n^2~G_{200}(e)~\sin 2 \eta ~=~ 0~~~,
   \label{*}
   \label{655}
   \ea
 an equation describing the evolution of the rotation angle $\,\eta\,$. As pointed out by Goldreich (1966), this equation is equivalent
 to the one describing  a simple pendulum. Indeed, in terms of $\,\beta=2\eta\,$, equation (\ref{655}) becomes $~\ddot{\beta}+
 \chi^2_{\textstyle{_{lib-max}}}\,\sin\beta=0~$, with a  constant positive $\,\chi^2_{\textstyle{_{lib-max}}}\,$ and with $\beta$ playing
 the role of the pendulum angle. \footnote{~In subsection \ref{libra} below, we write down the expression for the libration frequency
 $\,\chi_{\textstyle{_{lib-max}}}\,$ and also explain the reason why we equip it with such a subscript -- see formulae (\ref{665}) and
 (\ref{666}).}

 It follows directly from (\ref{655}) that the spin acceleration $\,\etadotdot\,$ vanishes if $\,\eta\,$ assumes the values of $\,0\,$ or
 $\,\pi\,$. As can be seen from the pendulum analogy, the initial conditions $\,(\eta\,,~\etadot\,)_{\textstyle{_{\,t=t_{\textstyle{_0}}}
 }}\,=\,(0\,,~0)\,$, as well as the conditions $\,(\eta\,,~\etadot\,)_{\textstyle{_{\,t=t_{\textstyle{_0}}}}}\,=\,(\pi\,,~0)\,$,
 correspond to the situation where the pendulum comes to a stall in the lower point (so $\,\betadot=0\,$ when $\beta=0$ or $2\pi$). Under
 such initial conditions, $\,\eta\,$ stays $\,0\,$ or $\pi$ all the time, which implies a uniform synchronous rotation of the secondary
 about the primary.~\footnote{~According to (\ref{655}), the spin acceleration $\,\etadotdot\,$ vanishes also for $\,\eta=\pm\pi/2\,$,
 which is the upper point of the pendulum.}

 Multiplication of equation (\ref{655}) by $\,\dot{\eta}\,$, with subsequent integration over time $\,t\,$, gives the first
 integral of motion,
   \ba
   \frac{1}{2} ~C~\etadot^2 ~ - ~ \frac{3}{4}~(B\,-\,A)~\frac{M_{sec}}{M_{sec}\,+\,M_{prim}} ~n^2 ~G_{200}(e)~\cos 2 \eta ~=~ E~~~,
   \label{**}
   \label{656}
   \ea
 whose value depends on the initial conditions.

 The period of the variable $\,\eta\,$ is given by a quadrupled integral over a quarter-cycle of $\,\eta~$:
 \ba
 P~=~4\,\int_{\eta=0}^{\eta=\eta_{\textstyle{_{max}}}}\;\frac{d\eta}{\etadot}~~~.
 \label{nonumber}
 \ea
 For circulation, use $\,\eta_{\textstyle{_{max}}}=\pi/2\,$. In the case of libration about $\,\eta=0\,$, the value of $\,\eta_{\textstyle{_{max}}}\,$ is less than $\,\pi/2\,$ and can be expressed via $\,E\,$ by setting $\,\etadot =0\,$ in equation (\ref{656}). To get an explicit
 expression of $\,P\,$, one should first express $\,\etadot\,$ via $\,E\,$ using (\ref{656}) and taking the positive root, and
 then should plug the so-obtained expression for $\,\etadot\,$ into (\ref{nonumber}). Also be mindful that libration about $\,\pi\,$
 can be converted to the same integral by adding or subtracting $\,\pi\,$ from the variable of integration.

 In what follows, averaging over the period $\,P\,$ will be denoted by $\,\langle...\rangle_{\textstyle{_{P}}}~$, the subscript serving to
 distinguish the operation from averaging over the orbit employed in Section \ref{ek} and in formula (\ref{z}).

 \subsection{Parallels with pendulum. The auxiliary quantity $\,W$}

 Goldreich (1966) noted the similarity of the differential equations (\ref{655} - \ref{656}) to the classical pendulum problem. If the
 initial conditions place the body outside of the 1:1 spin-orbit resonance, then $\,\eta\,$ circulates with a forced oscillation in
 rotation that depends on the mean value $\,\langle\etadot\rangle_{{_{P}}}\,$ of the $\,\etadot\,$ frequency. If however the initial
 conditions place the body's spin within a sufficiently close proximity of the 1:1 resonance (the ``resonance region"), then $\,\eta\,$
 will librate about $\,0\,$ or $\,\pi\,$. This will be a free physical libration with an amplitude smaller than $\,\pi/2\,$. Inside the
 resonance region, the initial conditions establish the free-libration amplitude and phase. However the restricted nature of the motion
 will keep the mean value of $\,\eta\,$ constant: it will be either $\,0\,$ or $\,\pi\,$. For the same reason, $\,\langle\etadot\rangle
 _{{_{P}}}\,$ will remain zero in the libration regime. In contrast to this, outside of the resonance region the mean rate of circulation
 $\,\langle\etadot\rangle_{{_{P}}}\,$ will possess a value determined by the initial conditions on $\,\eta\,$ and $\,\etadot\,$.

 Employing (\ref{656}) to express $\,\etadot\,$ via $\,E\,$, and plugging this expression into (\ref{nonumber}), Goldreich
 (1966) demonstrated that the period $\,P\,$ can be expressed via $\,E\,$, inside or outside the resonance region, by a
 complete elliptic integral of the first kind. This is natural, as the expression of $\,\etadot\,$ through $\,E\,$ involves
 a square root.

 Just as in the pendulum case, there exists a critical $\,E\,$ for which the period diverges. This is the boundary $\,E\,=\,E_b\,$
 separating circulation from libration. To locate the boundary, one should set simultaneously $\,\eta=\pi/2\,$ and $\,\etadot=0~$ in
 formula (\ref{656}):
 \ba
 E_b ~=~ \frac{3}{4}~(B\,-\,A)~\frac{M_{sec}}{M_{sec}\,+\,M_{prim}}~G_{200}(e)~n^2~~~.
 \label{657}
 \ea
 Mathematically, the emergence of a logarithmic singularity in $\,P\,$ at $\,E\,=\,E_b\,$ can be observed from formulae (9 - 10) in
 {\it{Ibid}}. Physically, this situation resembles the slowing-down of a pendulum at the circulation/libration border. In our
 problem, though, this division corresponds to $\,\eta=\pm\pi/2\,$.

 By setting simultaneously $\,\eta=0\,$ and $\,\etadot=0\,$ in (\ref{656}), we obtain the minimal value that the integral $\,E\,$ can
 assume:
 \ba
 \mbox{min}\,E~=~-~E_b~~~.
 \label{658}
 \ea
 The values $\,E\,>\,E_b\,$ correspond to circulation in either direction, with oscillations of the rotation rate. The values
 falling within the interval $\,E_b\,>\,E\,>\,-\,E_b\,$ give libration. The minimum value $\,E\,=\,-\,E_b\,$ corresponds to
 synchronous rotation without libration.

 The quantity $\,\eta\,$ being slowly varying (compared to the mean motion $\,n\,$), the period $\,P\,$, with which $\,\eta\,$ is
 changing, naturally turns out to be much longer than the orbital period, both outside and inside the 1:1 resonance:
 \ba
 P\,\gg\,\frac{2\,\pi}{n}~~~.
 \label{659}
 \ea
 This can be appreciated from the evident equalities
 \begin{subequations}
 \ba
 P~=~\frac{2~\pi}{\,|\,\langle\etadot\rangle_{{_{P}}}\,|\,}\quad\quad\quad\mbox{outside~of~the~1:1~resonance}~,
 \label{660a}
 \ea
 \ba
 P~=~\frac{2~\pi}{\chi_{\textstyle{_{lib}}}}\quad~\quad~\quad~\mbox{inside~the~1:1~resonance}~.\quad~
 \label{660b}
 \ea
 \label{660}
 \end{subequations}
 In (\ref{660b}), $\,\chi_{\textstyle{_{lib}}}\,$ denotes the libration frequency which too is much smaller
 than the mean motion $\,n\,$, as we shall see shortly.

 A useful quantity defined by Goldreich (1966) was
 \ba
 W~\equiv~P~\langle\etadot^2\rangle_{{_{P}}}~\equiv~\int_{0}^{P}\etadot^2~dt~=~4~\int_{\eta=0}^{\eta=\eta_{\textstyle{_{max}}}}\;\etadot~d\eta~~~.
 \label{661}
 \ea
 Similarly to (\ref{nonumber}), $\,\eta_{\textstyle{_{max}}}=\pi/2\,$ for circulation, and $\,\eta_{\textstyle{_{max}}}<\pi/2\,$ for libration.
 Just as $\,P\,$, so $\,W\,$ can be expressed through $\,E\,$ by complete elliptic integrals. For circulating $\,\langle\etadot\rangle_{{_{P}}}\,$,
 Goldreich described $\,P\,$ and $\,W\,$ as applying to one oscillation, but they describe one circulation of $\,\eta\,$ with two oscillations.
 For librations, $\,P\,$ and $\,W\,$ describe one libration cycle.

 One more quantity of use in this problem will be the mean square variation of $\,\etadot\,$ about $\,\langle\etadot\rangle_{{_{P}}}\,$,
 given by $\,\langle\etadot^2\rangle_{{_{P}}} - \langle\etadot\rangle_{{_{P}}}^2\,$. Be mindful that the notation $\,\langle...\rangle_{{_{P}}}\,$ is employed to indicate averaging over a circulation cycle as well as that over a libration cycle.

 When the values of the mean motion $\,n\,$, the mass factor $\,{\textstyle M_{sec}}/({\textstyle M_{sec}\,+\,M_{prim}})\,$, and
 the ratio $\,(B-A)/C\,$ are given, and the initial conditions on $\,\eta\,$ and $\,\etadot\,$ are set,
 equation (\ref{656}) furnishes the value of $\,E/C\,$. The comparison of this value with $\,E_b/C\,$ distinguishes circulation
 from libration. With the values of $\,(B-A)/C\,$ and $\,E/C\,$ known, the complete elliptic integrals can be evaluated and the
 values of the quantities $\,P\,$, $\,W\,$, $\,\langle\etadot\rangle_{{_{P}}}\,$, and $\,\langle\etadot^2\rangle_{{_{P}}}\,$ can be found. It
 should also be possible to reverse this procedure. For example, the knowledge of the circulating value of $\,\langle\etadot\rangle_{{_{P}}}
 \,$, along with the knowledge of $\,n\,$, the mass factor $\,{\textstyle M_{sec}}/({\textstyle M_{sec}\,+\,M_{prim}})\,$, and
 the ratio $\,(B-A)/C\,$, should allow $\,E/C\,$, $\,W\,$, and $\,\langle\etadot^2\rangle_{{_{P}}}\,$ to be computed.

 \subsection{The tidal torque and the libration bias}

 To account for tidal dissipation, Goldreich (1966) added the
 averaged (over an orbital period) tidal torque (\ref{geg}) to the right-hand side of equation (\ref{655}). We however shall
 employ the corrected average torque (\ref{Laskar_av}) instead of (\ref{geg}). This will result in
 \ba
 \nonumber
 C\etadotdot+\,\frac{3}{2}(B-A)\,\frac{M_{sec}}{M_{sec}+M_{prim}}~n^2\,G_{200}(e)~\sin 2\eta~=~\quad\quad\quad\quad~~~~\quad~~~~ ~\quad\quad\quad\quad~~~~ ~\quad\quad\quad\quad~~~~ \\
 \nonumber\\
 ~-~{\cal{Z}}\left[\etadot
 \left(1+\,\frac{\textstyle 15}{\textstyle 2}~e^2\right)\,-\,6\,n\,e^2\left(1+\,\frac{121}{16}\;e^2\right)\right]
 ~+~O(e^6)~+~O(\inc^2/Q)~+~
 O(en\Delta t/Q)\,~,~~~~\quad~~
 \label{***}
 \label{662}
 \ea
 where the coefficient $\,{\cal{Z}}\,$ depends via (\ref{54}) on the orbital variables, the tidal parameters, and the masses.

 As we already mentioned at the beginning of Section \ref{ek}, ignoring the permanent-figure term
 would lead one to the conclusions that the tidal torque is finite for synchronous rotation and that it vanishes for a spin rate
 slightly higher than synchronous. However inclusion of the permanent-figure term into the picture alters the results radically. The
 constant term on the right-hand side of equation (\ref{***}) causes the mean value of $\,\eta\,$ to be slightly larger than $\,0\,$ or
 $\,\pi\,$. For small librations, this bias is
 \begin{subequations}
 \ba
 \eta_{\textstyle{_{bias}}}&=&\frac{2\;{\cal{Z}}\;e^2}{(B-A)\,n}~\frac{M_{sec}+M_{prim}}{M_{sec}}\left(1+\,
 \frac{121}{16}\,e^2\right)\,\frac{~~1}{G_{200}(e)}\,+\,O(e^6)\,+\,O(\inc^2/Q)\,+\,
 O(en\Delta t/Q)~\quad~\,~\quad~\,\quad\\
 \nonumber\\
 \nonumber\\
 &=&\frac{2~{\cal{Z}}~e^2}{(B-A)\,n}~\frac{M_{sec}+M_{prim}}{M_{sec}}\left(1+\,
 \frac{161}{16}\,e^2\right)\,+\,O(e^6)\,+\,O(\inc^2/Q)\,+\,
 O(en\Delta t/Q)\,~,~\,\quad~\quad~\,\quad
 \ea
 \label{669}
 \end{subequations}
 with $\,{\cal{Z}}\,$ calculated through (\ref{54}). The value of the time lag $\,\Delta t\,$
 entering the expression (\ref{54}) for $\,{\cal{Z}}\,$ should be set appropriate to the
 orbital frequency (mean motion).

 As evident from (\ref{669}), the bias comes into being due to the eccentric shape of the
 orbit. Bodies with permanent figures can achieve synchronous rotation because the bias in
 $\,\eta\,$ gives birth to a permanent-figure torque that balances the constant dissipative
 torque.

 At this point, an important caveat will be in order. As we explained in Section \ref{ger}, the MacDonald model of tides
 becomes self-consistent only when the time delay $\,\Delta t\,$ emerging in (\ref{54}) is set frequency-independent. This
 circumstance limits the precision of the MacDonald torque and of the Goldreich dynamical model based thereon,
 whenever the model gets employed to determine the timescales of spin evolution (Efroimsky \& Lainey 2007). Nevertheless, for very
 slow evolution the model may be employed for obtaining qualitative estimates. In this case, $\,\Delta t\,$ should be treated as a
 parameter that itself depends upon the forcing frequency in the material.
 For rotation outside the $\,1:1\,$ spin-orbit lock, it would be a tolerable approximation to use $\,\Delta t\,$ appropriate to the
 principal tidal frequency $\,\chi_{\textstyle{_{2200}}}\,$ or to the double instantaneous synodic frequency (\ref{duga}). However
 inside the $\,1:1\,$ resonance, $\,\Delta t\,$ would correspond to the libration frequency $\,\chi_{\textstyle{_{lib}}}\,$ which
 may be {\it{very}}$\,$ different from the usual tidal frequencies for nonsynchronous rotation. This circumstance may change the
 value of $\,\Delta t\,$ and therefore of $\,{\cal{Z}}\,$ noticeably.

 \subsection{Tidal dissipation and the point of stall}

 Be mindful that on the right-hand side of equation (\ref{662}) we have the tidal torque {\it{averaged over the orbit}} (see Appendix B
 for details). Similarly, the second term on the left-hand side of (\ref{662}) is the permanent-figure torque {\it{averaged over the
 orbit}} -- see expression (\ref{x}).\footnote{~For the first term, $\,C\ddot{\eta}\,$, the caveat about orbital averaging is unimportant,
 because $\,\etadot\,$ bears no dependence upon the mean anomaly.} Therefore equation (\ref{662}) renders us the behaviour
 of $\,\eta\,$ over times longer than the orbital period. This is acceptable, because the orbital period is much shorter than the
 timescales of our interest. Specifically, it is much shorter than $\,P\,$, see equation (\ref{659}).

 Without dissipation, $\,E\,$ is conserved during oscillations of $\,\eta\,$, while in the presence of weak dissipation $\,E\,$ changes slowly. Indeed,
 multiplying both sides of (\ref{662}) by $\,\etadot\,$ and making use of (\ref{656}), we arrive at
   \ba
   \frac{dE}{dt}\,=\,-\,{\cal{Z}}\,\left[\,\etadot^2\,\left(\,1\,+\,\frac{15}{2}\,e^2\,\right)\,
   -\,6\,n\,\etadot\,e^2\,\left(\,1\,+\,\frac{121}{16}\,e^2\,\right)\,\right]\,+\,O(e^6)\,+
   \,O(\inc^2/Q)\,+\,O(en\Delta t/Q)~~.~~~
   \label{****}
   \label{663}
   \ea
 Thus we see that a dissipative tidal torque influences the spin, while causing changes also in the first integral $\,E\,$ (which is {\it{not}}
 identical to the actual kinetic energy). Considering equation (\ref{656}) and noting that the cosine term is periodic, we see
 that weak tidal dissipation will cause a change in $\,\etadot^2\,$ over time scales long compared to $\,P\,$.

 For nonsynchronous rotation, weak dissipation causes both a slow secular change and a small oscillation. The latter arises from the oscillating part
 of $\,\etadot\,$ as $\,\eta\,$ circulates or librates.

 Goldreich (1966) also averages over the period $\,P\,$. For libration, the so-averaged rate
 of change of $\,E\,$ is
   \ba
   \langle\,{d E  }/{dt}\,\rangle_{{_{P}}}~=~-~\frac{{\cal{Z}}~W}{P}~\left(\,1~+~\frac{15}{2}~e^2\,\right)
   \,+\,O(e^6)\,+\,O(\inc^2/Q)\,+\,O(en\Delta t/Q)~~~,
   \label{664}
   \ea
 where the positive definite quantity $\,W\,$ is introduced via (\ref{661}). The negative $~\langle\,{d E  }/{dt}\,\rangle_{{_{P}}}~$ for libration means that the maximum $\,\etadot^2\,$ at $\,\cos 2\eta = 1\,$ in equation (\ref{**}) is decreasing
 and the free libration damps with time. Libration evolves toward synchronous rotation with $\,\etadot = 0\,$ and $\,\eta\,$ equal
 to $\,0\,$ or to $\,\pi\,$.

 The $\,\etadot\,$ term causes the damping of the libration amplitude. When the amplitude is sufficiently small, its decrease obeys
 the exponential law $\,\exp(-D_{\textstyle{_L}}\,t)\,$ with
 \ba
 D_{\textstyle{_L}}~=~\frac{\cal Z}{2\,C}~\left(\,1~+~\frac{15}{2}~e^2\,\right)~~~.
 \label{668}
 \ea
 As we emphasised in the paragraph after equation (\ref{662}), inside the $\,1:1\,$ resonance the time delay $\,\Delta t\,$ showing
 up in the expression for $\,{\cal Z}\,$ should be appropriate to the libration frequency $\,\chi_{\textstyle{_{lib}}}\,$ which
 may differ greatly from the usual tidal frequencies for nonsynchronous spin. This choice of $\,\Delta t\,$ will influence the
 value of $\,{\cal{Z}}\,$.

 For circulation, averaging of equation (\ref{663}) over one orbital period leads to
   \ba
   \nonumber
   \langle\,{d E  }/{dt}\,\rangle_{{_{P}}}~=&-&\frac{\cal Z}{P}~\left[\,W~\left(1~+~\frac{15}{2}~e^2\right)~-~12~\pi~n~e^2~\left(\,1\;+\;
   \frac{121}{16}\;e^2\,\right)~\mbox{sgn}\,\langle\etadot\rangle_{{_{P}}}\,\right]~~~,\\
   \nonumber\\
   \nonumber\\
   &&+\,O(e^6)\,+\,O(\inc^2/Q)\,+\,O(en\Delta t/Q)
   \label{xxxb}
   \label{670}
   \ea
 where we have used definition (\ref{660a}) for $\,P\,$ and definition (\ref{661}) for $\,W\,$.

 While $\,W = P \langle\etadot^2\rangle_{{_{P}}}\,$ is positive definite, the second term on
 the right-hand side of equation (\ref{670}) can, for circulation, have either sign. In the
 case when its sign is positive, the expression (\ref{670}) for $\,\langle\,{d E  }/{dt}\,
 \rangle_{{_{P}}}\,$ will vanish for $\,W\,$ equal to
   \ba
   W_{\textstyle{_{stall}}}~=~12~\pi~n~{e^2}\left(1\;+\;\frac{\textstyle e^2}{\textstyle 16}\,\right)
   \,+\,O(e^6)\,+\,O(\inc^2/Q)\,+\,O(en\Delta t/Q)~~~.
   \label{672}
   \ea
 No matter whether $\,W\,$ begins its evolution with an initial value larger or smaller than $\,W_{\textstyle{_{stall}}}\,$, it
 will never cross $\,W_{\textstyle{_{stall}}}~$.

 Another important value of $\,W\,$ is the one corresponding to the boundary between libration and circulation, $\,W_b\,$. Below we
 shall obtain its value and shall explain that for $\,W_{b}<W_{\textstyle{_{stall}}}\,$ there exists a positive value of
 $\,\langle\etadot\rangle_{{_{P}}}\,$ at which the evolution of a circulating $\,\langle\etadot\rangle_{{_{P}}}\,$ should stall.

 \subsection{The free-libration frequency}\label{libra}

 Exploring librations, we start out with the small-amplitude case. Replacing $\,\sin 2 \eta\,$ with $\,2\eta\,$ in
 equation (\ref{*}), we write down the frequency for {\it{small}} librations of $\,\eta~$:
 \ba
 \chi_{\textstyle{_{lib-max}}}~=~\frac{2\,\pi}{P_{min}}~=~\left[\,3~\frac{B-A}{C}~\,\frac{M_{sec}}{\,M_{sec}\,+\,M_{prim}\,}~G_{200}(e)\,\right]^{
 1/2}n~~~.
 \label{665}
 \ea
 In many realistic situations, the condition $\,\frac{\textstyle B-A}{\textstyle C}~\frac{\textstyle M_{sec}}{\textstyle M_{sec}\,+\,
 M_{prim}}\,\ll\,1~$ is fulfilled,\footnote{~This condition is satisfied safely if the primary is a planet (like Mercury) or a large
 satellite (like our Moon). Its fulfilment is not guaranteed, though, for small satellites (like Phobos or Hyperion).} which ensures
 that $\,\chi_{\textstyle{_{lib-max}}}\ll n\,$. Larger amplitudes increase the period $\,P\,$ and render a smaller frequency
 $\,\chi_{\textstyle{_{lib}}}\,$, so expression (\ref{665}) gives the smallest librating period and the {\it{largest}}$\,$ frequency
 \ba
 \chi_{\textstyle{_{lib-max}}}\,=\,\mbox{max}\,\chi_{\textstyle{_{lib}}}~~~.
 \label{666}
 \ea
 The linear $\,\etadot\,$ term in (\ref{***}) will alter the frequency, but for slow tide-caused evolution the correction will be small,
 so (\ref{665}) still will serve well as an approximation for the maximal frequency of libration. On all these grounds, we now accept
 that for both small-amplitude and large-amplitude librations the inequality
 \ba
 \chi_{\textstyle{_{lib}}}~\ll~n
 \label{667}
 \ea
 holds. This justifies, {\it{a posteriori}}, the assertion made after (\ref{660b}).

 In the literature on the physical libration of the Moon, the expression for the free-libration frequency is ubiquitous, though often
 without the mass factor. Versions of the expressions for $\,\eta_{\textstyle{_{bias}}}\,$ and the free-libration damping rate appeared
 in Williams et al. (2001). That paper, though, did not rely on the corrected MacDonald model (constant time delay), but simply used
 separate $\,Qs\,$ for libration and orbital frequencies.

 \subsection{The boundary between circulation and libration}

 The boundary between circulation and the resonance zone corresponds to a value $\,W\,=\,W_b\,$. To find it, we combine (\ref{656}) with (\ref{657})
 and arrive at
 \begin{subequations}
 \ba
 \frac{1}{2}\;C\;\etadot^2\,=\;E_b\,\left(\,1\,+~\cos2\eta\,\right)~~~,
 \label{}
 \ea
 which is the same as
 \ba
 \etadot^2\,=\;4\;\frac{E_b}{C}\,\cos^2\eta~~~.
 \label{}
 \ea
 \end{subequations}
 Insertion of the resulting expression for $\,\etadot\,$ into (\ref{661}) entails:
 \ba
 W_b~=~8\,\sqrt{\frac{E_b}{C}}\,\int_{0}^{\pi/2}\cos\eta~d\eta
 ~=~8~\sqrt{\frac{E_b}{C}}~=~ 4 ~n~ \left[\,3~\frac{B-A}{C}~\frac{M_{sec}}{\,M_{sec}\,+\,M_{prim}\,}~G_{200}(e)\,\right]^{1/2} ~~,~~~
 \label{671}
 \ea
 comparison whereof to (\ref{665}) yields another expression for the boundary value:
 \ba
 W_b = 4~ \chi_{\textstyle{_{lib-max}}}~~~.
 \nonumber
 \ea
 While $\,W\,$ is continuous across the boundary, $\,P\,$ has a logarithmic singularity, as was mentioned in the paragraph
 after equation (\ref{657}). As $\,P\,$ diverges, the evolution rate of $\,\langle\,{d E  }/{dt}\,\rangle_{{_{P}}}\,$, given by (\ref{670}) , vanishes at the boundary.
 Nonetheless a small perturbation allows the boundary to be crossed -- for more on this, see the two paragraphs after formula (22)
 in Goldreich (1966).

 \subsection{Three regimes}

 If $\,W_{\textstyle{_{stall}}}\,<\,W_b\,$, then there is no stall point in the
 region of circulation ($\,W_b\,<\,W\,$). So $\,W\,$ evolves towards $\,W_b\,$, while $\,\langle\etadot\rangle_{{_{P}}}\,$ of either sign
 evolves towards zero at the libration/circulation boundary. Figure 1a illustrates the evolution of $\,\langle\etadot\rangle_{{_{P}}}\,$.
 Goldreich (1966) comments that the boundary will be crossed, the free libration will damp, and the rotation will evolve toward the
 synchronous state. The synchronous state has a zero $\,\etadot\,$, with $\,\eta\,$ biased slightly off$\,$\footnote{~The value
 $\,\pi\,$ shows up because of the factor 2 accompanying $\,\eta\,$ when this variable enters $\,\sin 2\eta\,$ in the equation for
 the torque and $\,\cos 2\eta\,$ in the expression $\,E\,$. The tiny bias off $\,\eta=0\,$ or off $\,\eta=\pi\,$, given by (\ref{669}), will emerge
 due to dissipation.} of either $\,0\,$ or $\,\pi\,$. The figure does not show the libration region.\footnote{~A figure with $\,W\,$
 rather than $\,\langle\etadot\rangle_{{_{P}}}\,$ would show the libration region, along with the circulating region. On such a figure,
 though, positive and negative values of $\,\langle\etadot\rangle_{{_{P}}}\,$ would overlap, so that the one-sided nature of the stall would
 not be evident.}

 \begin{figure}[h]
  \hspace{0.5cm}
  {\includegraphics[width=14.1cm]{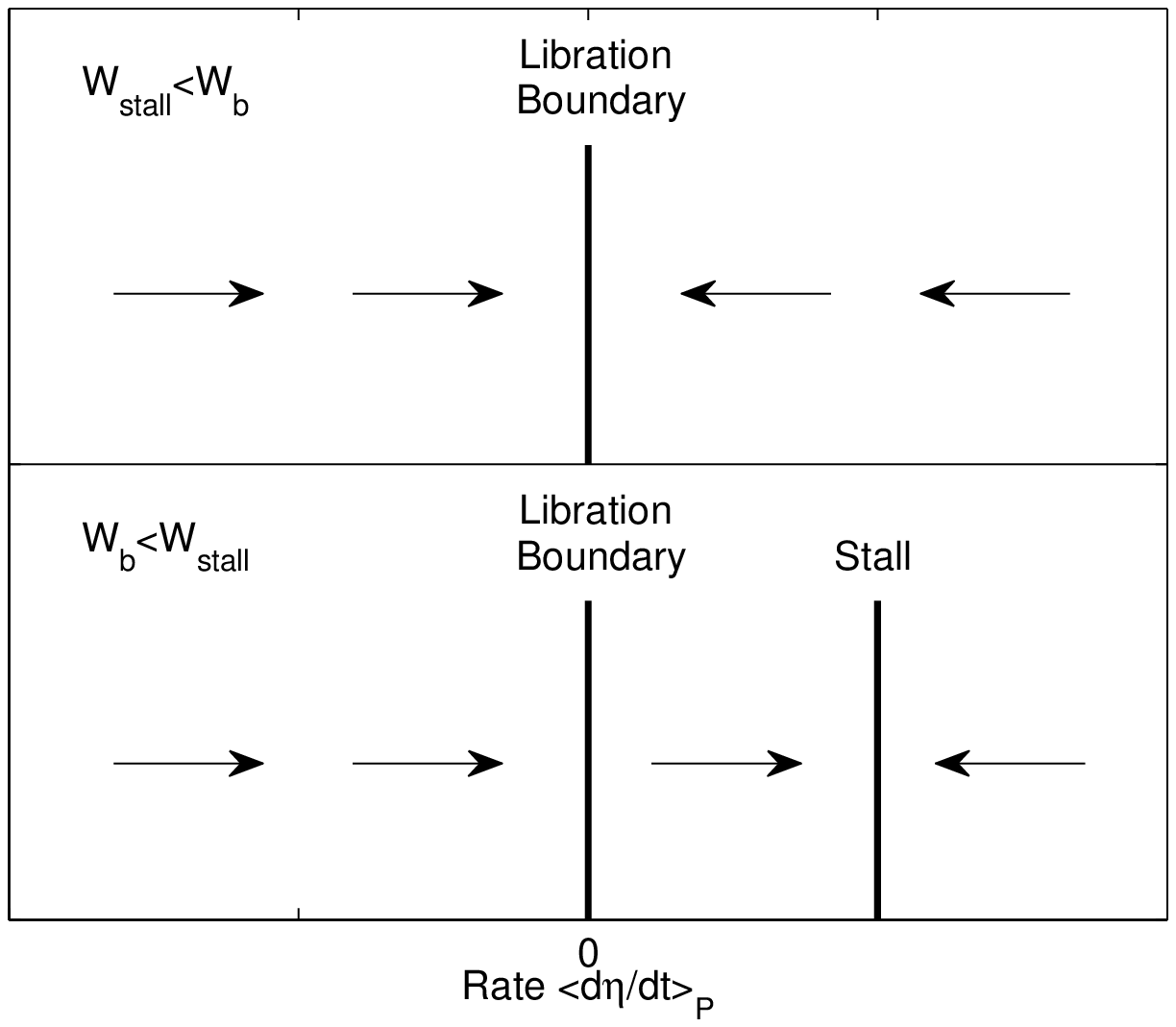}}
 {\caption{\small{Three possible scenarios of evolution of $\,\langle\etadot\rangle_{{_{P}}}\,$.
 Arrows on the top diagramme illustrate the evolution of circulating $\,\langle\etadot\rangle_{{_{P}}}\,$ toward the
 libration boundary for $\,W_{\textstyle{_{stall}}}\,<\,W_b\,$.
 Arrows on the left of the bottom diagramme depict the evolution of negative $\,\langle\etadot\rangle_{{_{P}}}\,$ toward the libration
 boundary. Arrows in the midst and on the right of the bottom diagramme show the evolution of positive $\,\langle\etadot\rangle_{{_{P}}}\,$
 toward the stall point for $\,W_b\,<\,W_{\textstyle{_{stall}}}\,$.
 }}}
 \label{Figure_1}
                                                                                 \vspace{6.0cm}
 \end{figure}

 ~\\
 If $\,W_b\,<\,W_{\textstyle{_{stall}}}\,$, then a stall point exists in the circulating region ($\,W_b\,<\,W\,$).
  All in all, Figure 1b summarises the following three cases:\\
 \begin{itemize}

\item[\bf A.~] For an initially negative $\,{\langle\etadot\rangle}_{{_{P}}}\,$, circulation is taking place. We have $\,W_{b}<W\,$,
 but the stall point is never located on the negative $\langle\etadot\rangle_{{_{P}}}\,$ side of zero. The rate
 $\,\langle\,dE/dt\,\rangle_{{_{P}}}\,$ is negative, as can be seen from (\ref{663}). We then have a
 slow decrease of the three quantities:\footnote{~Be mindful that $\,E\,$ remains a constant over a cycle of $\,P\,$, except
 for a tiny amount of dissipation during a cycle.} $\,E\,\rightarrow\,E_b~$, $~W\,\rightarrow\,W_b\,$, and $\,\langle\etadot\rangle_{{_{P}}}
 \,\rightarrow\,0\,$. This decrease takes the system to the circulation/libration boundary. Goldreich (1966) commented that it
 is ambiguous whether $\,\langle\etadot\rangle_{{_{P}}}\,$ would cross zero and proceed to increase or whether the boundary would be
 crossed passing into the libration region followed by damping of the libration and by evolution toward the synchronous state.

 Be mindful that, while the first integral $\,E\,$ is decreasing in the circulation case, the actual kinetic energy of rotation
 is increasing. Indeed, when the negative $\,\langle\etadot\rangle_{{_P}}\,$ is evolving towards zero, the spin rate $\,\thetadot\,$ is growing, and
 so is the rotational energy $\,C\thetadot^2/2\,$.\\

\item[\bf B.~] For an initially positive $\,\langle\etadot\rangle_{{_{P}}}\,$ lying between the circulation/libration boundary and the
 stall point, i.e., for $\,W_b\,<\,W\,<\,W_{\textstyle{_{stall}}}\,$, the rate $\,\langle\,dE/dt\,\rangle_{{_{P}}}\,$ is positive, and
 $\,E\,$ increases. The quantities $\,W\,$, $\,\langle\etadot^{\,2}\rangle_{{_{P}}}\,$ and $\,\langle\etadot\rangle_{{_{P}}}\,$ will evolve to larger
 values until the evolution stalls as $\,W\,$ approaches $\,W_{\textstyle{_{stall}}}\,$ from below.\\

 As the positive $\,\etadot\,$ is increasing, so are the spin rate $\,\thetadot\,$ and the rotational energy.\footnote{~In the first
 integral (\ref{656}), the kinetic-energy-like part is given by $\,C \etadot^2/2=(C/2)(\thetadot-n)^2=(C/2)(\thetadot^2-2n\thetadot +
 n^2)~$. When the spin accelerates, the additional kinetic energy is borrowed from the orbital motion. Because of the dissipation, the
 overall spin + orbit energy must nevertheless decrease. (The orbital energy includes a kinetic part and the $\,-GM/r\,$ potential
 term.)}

\item[\bf C.~] For an initially positive $\,\langle\etadot\rangle_{{_P}}\,$ beyond the stall point we have: $\,W_b\,<\,W_{\textstyle{_{stall}}}
 \,<\,W\,$. Then $\,\langle\,dE/dt\,\rangle_{{_{P}}}\,$ is negative and $\,E\,$ is slowly decreasing.\footnote{~Just as in
 item {\bf{A}}, here $\,E\,$ stays virtually unchanged over $\,P\,$.}
 Likewise, $\,W\,$, $\,\langle\etadot^{\,2}\rangle_{{_{P}}}\,$ and $\,\langle\etadot\rangle_{{_{P}}}\,$ evolve to lower values until the
 evolution stalls as $\,W\,$ approaches $\,W_{\textstyle{_{stall}}}\,$ from above.

 The decrease of the positive $\,\etadot\,$ renders a decrease in the rotation rate $\,\thetadot\,$ and thus leads to damping of
 the kinetic energy of rotation.\\

 \end{itemize}

 Calculation of the frequency at the stall point $\,\langle\etadot_{\textstyle{_{stall}}}\,\rangle_{_{P}}\,$ would require two steps:
 first, reversing the elliptic integral computation, starting with $\,W_{stall}\,$; and second, deriving $\,E/C\,$, $\,P_{stall}\,$
 and $\,\langle\etadot_{\textstyle{_{stall}}}\,\rangle_{{_{P}}}~$.

 At the stall point, formulae (\ref{660a}) and (\ref{661}) acquire the form of
 \ba
 P_{\textstyle{_{stall}}}\,=~\frac{2~\pi~}{~\langle\etadot_{\textstyle{_{stall}}}\rangle_{{_{P}}}}
 \quad~\quad
 \mbox{and}
 \quad~\quad
 W_{\textstyle{_{stall}}}\,=~P_{\textstyle{_{stall}}}~\langle\etadot^{\,2}_{\textstyle{_{stall}}}\,\rangle_{{_{P}}}~~~,
 \label{}
 \ea
 correspondingly. In combination with (\ref{672}), this entails:
 \ba
 \frac{\langle\dot{\eta}_{\textstyle{_{stall}}}^{\,2}\rangle_{{_{P}}}}{\langle\dot{\eta}_{\textstyle{_{stall}}}\rangle_{{_{P}}}}~=~{6\,n\,e^2}
 \left(1\;+\;\frac{\textstyle e^2}{\textstyle 16}\,\right)~~~,
 \label{b}
 \ea
 whence we once again note that the frequency $\,\langle\etadot_{\textstyle{_{stall}}}\,\rangle_{{_{P}}}\,$ must be positive. The comparison of $\,W_{\textstyle
 {_{stall}}}\,$ with $\,W_b\,$ in the inequalities mentioned in the above items {\bf{A}}, {\bf{B}}, {\bf{C}} is then equivalent to comparing $\,\langle
 \etadot^2_{\textstyle{_{stall}}}\,\rangle_{{_{P}}} / \langle\etadot_{\textstyle{_{stall}}}\rangle_{{_{P}}}\,$ with $\,(2/\pi)\,\chi_{\textstyle{_{lib-max}}}\,$ or to
 comparing\footnote{The latter may also be expressed as comparison of $\,3\pi e^2\left(1+\frac{\textstyle 21}{\textstyle 16}\,e^2\right)\,$
 with $\,\left[3\,\frac{\textstyle B-A}{\textstyle C}~\frac{\textstyle M_{Earth}}{\textstyle M_{Earth}+M_{Moon}}\right]^{1/2}\,$.} $\,3
 \,\pi\,e^2\,(1+e^2/16 )\,$ with $\,\left[3~\frac{\textstyle B-A}{\textstyle C}~\frac{\textstyle M_{Earth}}{\textstyle M_{Earth}+M_{
 Moon}}~G_{200}(e)~\right]^{1/2}\,$.

 A comment on equation (\ref{b}) would be in order. Since oscillations of $\,\eta\,$ result from the existence of the permanent
 triaxiality, evolution of $\,\eta\,$ becomes smooth in the $\,A\,=\,B\,$ limit. Averaging becomes unnecessary, so $~\langle\dot{
 \eta}_{\textstyle{_{stall}}}^{\,2}\rangle_{{_{P}}}~$ becomes simply $~\dot{\eta}_{\textstyle{_{stall}}}^{\,2}~$, while $\,\langle\dot{\eta
 }_{\textstyle{_{stall}}}\rangle_{{_{P}}}\,$ becomes $\,\dot{\eta}_{\textstyle{_{stall}}}\,$. This way, in the oblate-body case considered
 back in Section \ref{ek}, equation (\ref{b}) acquires the form of $~\etadot_{\textstyle{_{stall}}}=\,{6\,n\,e^2}
 \left(1\;+\;{\textstyle e^2}/{\textstyle 16}\,\right)~$, which agrees with (\ref{kik}).

 \subsection{Application to the Moon}

 When Goldreich (1966) applied his inequality expressions to the Moon, he found that the stall point would have interrupted evolution of
 rotation from faster spin to synchronous rotation. Here we have repeated his study, though with the corrected average torque
 (\ref{Laskar_av}) instead of (\ref{geg}). $~$For $~\frac{\textstyle B-A}{\textstyle C}\,=\,2.278\,\times\,10^{-4}\,$ (Williams \&
 Boggs 2012)$\,$, $~e=0.0549\,$, $\,$and $~\frac{\textstyle M_{Earth}}{\textstyle M_{Earth}+M_{Moon}} = 0.98785~$, $\,$the libration
 period turns out to be 38 times the orbital period, while $\,W_{stall}\,$ is $\,10\,\%\,$ larger than $\,W_{b}\,$. This renders a value
 of $\,W_{stall}/W_{b}\,$ much smaller than the one found by Goldreich (1966), and the difference is mainly due to our use of the
 corrected average torque (\ref{Laskar_av}). Despite the so-different value of $\,W_{stall}/W_{b}\,$, despinning of the Moon would still
 be interrupted by a stall. In principle, a tidal spin-up scenario remains an option too, though this option does not look probable.

 Goldreich (1966) noted that the lunar orbital eccentricity is changing. To make the stall point disappear, i.e., to ensure that
 $\,W_{stall}<W_b\,$. the eccentricity of the Moon would need to be less than $\,95\,\%\,$ of its present value. While we lack data on
 the ancient evolution of the lunar eccentricity, the modern eccentricity rate of about $2\times 10^{-11}\,$yr$^{-1}$ has been reliably
 determined through analysis of the Lunar Laser Ranging data (Williams et al. 2001, Williams \& Boggs 2009). For the measured
 eccentricity rate, the eccentricity would be small enough to prevent a stall prior to $\,1.4\times10^8\,$ yr ago. The Moon has clearly
 been a satellite of the Earth for billions of years, and the capture into the synchronous spin state should have occurred very early in
 its history.

 For the Moon as it exists today, the free libration in longitude has a $\,2.9\,$ yr period. The damping time is four orders of
 magnitude longer (Williams et al. 2001), so evolution of this libration is slow. Despite the damping time being short compared to the
 lunar age, the Moon has a small free libration amplitude of $\,1.3"\,$ (Rambaux and Williams 2011). There has been geologically recent
 stimulation, probably due to resonance crossing (Eckhardt 1993).

 The ambiguity of evolution of the negative $\,\langle\etadot\rangle_{{_{P}}}\,$ past the
 circulation/libration boundary deserves a comment. For Mercury, Makarov (2012) finds that a
 more realistic tidal dissipation model than the corrected MacDonald torque strongly changes
 computations of the evolution of planetary spin rate near the $\,3:2\,$ and higher
 spin-orbit resonances.

 \section{Conclusions}

 In the article thus far, we have provided a detailed explanation of how the empirical MacDonald model can be derived from a more
 accurate and comprehensive Darwin-Kaula theory of bodily tides. We have demonstrated that the derivation hinges on a key assertion
 that the quality factor $\,Q\,$ of the primary should be inversely proportional to the tidal frequency. This crucial circumstance
 was missed by MacDonald (1964), who made his theory inherently self-contradictory by setting the quality factor to be a
 frequency-independent constant.

 We have corrected this oversight in the MacDonald approach, and have developed an appropriate correction to Goldreich's model of
 spin dynamics and evolution near the 1:1 spin-orbit resonance. Although we got different numbers, qualitatively the main conclusion
 by Goldreich (1966) stays unaltered: when an oblate body's spin is evolving toward the resonance, vanishing of the average tidal torque
 still implies a pseudosynchronous rotation (rotation slightly faster than resonant), while synchronicity requires a small compensating torque. For a triaxial body, the picture gets
 more complex due to the emergence of a triaxiality-caused torque. (While the oblate case is appropriate for spinning gaseous or
 liquid planets and moons, the triaxial case applies to rocky objects.)

 Goldreich (1966) linked the possible trapping of a body in the synchronous state during its tidal evolution of rotation to its
 triaxiality. In light of the correction of expression (\ref{oldie}) to (\ref{kik}), that limiting condition between the triaxiality
 and $\,e^2\,$ must change. After capture into the synchronous state, the tidal torque is compensated by a triaxiality torque by
 aligning the principal axis associated with the smallest moment of inertia slightly off of the mean direction to the external body.
 A constant is thereby introduced into the physical libration in longitude.

 Setting the tidal quality factor to scale as inverse frequency is incompatible with the actual dissipative properties of realistic
 mantles and crusts. Nevertheless, after the afore-explained correction is implemented, both the corrected MacDonald description of
 tides and the dynamical theory based thereon remain valuable toy models capable of providing a good qualitative handle on tidal
 dynamics over not too long timescales. Specifically, this approach renders a simple qualitative description of the interplay
 between the tidal torque and the triaxiality-caused torque exerted on a body near the 1:1 spin-orbit resonance.

 It should also be remembered that the stability of pseudosynchronous rotation hinges upon the dissipation model employed. Makarov and Efroimsky (2013) have found that a more realistic tidal dissipation model than the corrected MacDonald torque makes pseudosynchronous rotation unstable. Finally, pseudosynchronism becomes impossible when the triaxiality is too large (see subsection 6.7, specifically footnote 24).

 {\Large{\bf Acknowledgements}}\\

 We are indebted to Sylvio Ferraz-Mello for numerous fruitful discussions on the theory of
 bodily tides and for referring us to the paper by Rodr{\'{\i}}guez, Ferraz-Mello \& Hussmann
 (2008). We also express our deep thanks to Anthony Dobrovolskis whose review of our
 manuscript was very comprehensive and extremely helpful.

 A portion of the research described in this paper was carried out at the Jet Propulsion Laboratory of the California Institute of
 Technology, under a contract with the National Aeronautics and Space Administration. Government sponsorship acknowledged.\\
 ~\\


 \noindent
 {\underline{\textbf{\Large{Appendices.}}}}

 \appendix

 \section{The tidal-torque vector\\ and its components in spherical coordinates}\label{tor}

 A secondary with spherical coordinates $\,\erbold^{\,*}\,=\,(r^{\,*},\,\lambda^*,\,\phi^*)\,$ and mass $\,M_{sec}^{\,*}\,$ raises a
 tidal bulge on the primary. The gravitational attraction between the tidal bulge and a secondary at $\,\erbold\,=\,(r\,,\,\lambda\,
 ,\,\phi)\,$ with mass $\,M_{sec}\,$ causes equal but opposite torques on the primary and the secondary. For the external tidal
 potential $\,U(\erbold)\,$, the torque components depend on the partial derivatives of the potential $\,U\,$ along great circle
 arcs.

 To calculate these expressions, let us recall some basics. The torque $\,\vec{\bf T}\,$ wherewith the primary is acting on the
 secondary is given by the cross-product
 \ba
 \vec{\bf T}~=~\erbold\times{\vec{\cal F}}~~~,
 \label{thereon}
 \ea
 $\vec{\cal F}\,$ being the tidal force exerted by the primary on the secondary. This force is given by
 \ba
 {\vec{\cal F}}~=~-~M_{sec}~\frac{\partial U(\erbold)}{\partial \erbold}~=~-~M_{sec}~\left(~\frac{\partial U}{\partial r}~\hat{\bf{e}}_{\textstyle{_{r}}}
 \,+~\frac{1}{r}~\frac{1}{\cos\phi}~\frac{\partial U}{\partial
 \lambda}~\hat{\bf{e}}_{\textstyle{_{\lambda}}}\,-~\frac{1}{r}~\frac{\partial U}{\partial \phi}~\hat{\bf{e}}_{\textstyle{_{\phi}}}~\right)~~~,
 \label{therewith}
 \ea
 $\hat{\bf{e}}_{\textstyle{_{r}}}\,,\,\hat{\bf{e}}_{\textstyle{_{\lambda}}}\,,\,\hat{\bf{e}}_{\textstyle{_{\phi}}}\,$ being the unit vectors of a spherical coordinate system associated
 with the primary's equator and corotating with it.\footnote{~Were we using the polar angle $\,\varphi=\pi/2-\phi\,$ instead of the
 latitude $\,\phi\,$, the right-handed triple of unit vectors would be: $\,\hat{\bf{e}}_{\textstyle{_{r}}}\,,\,\hat{\bf{e}}_{\textstyle{_{\varphi}}}\,,\,\hat{\bf{e}
 }_\lambda\,$ (``radial -- south -- east"), while the gradient would read as $~\frac{\textstyle\partial U(\erbold)}{\textstyle\partial
 \erbold}~=~\frac{\textstyle\partial U}{\textstyle\partial r}~\hat{\bf{e}}_{\textstyle{_{r}}}\,+~\frac{\textstyle 1}{\textstyle r}~
 \frac{\textstyle\partial
 U}{\textstyle\partial \varphi}~\hat{\bf{e}}_{\textstyle{_{\varphi}}}\,+~\frac{\textstyle 1}{\textstyle r}~\frac{\textstyle 1}{\textstyle \sin\varphi}~\frac{
 \textstyle \partial U}{\textstyle \partial \lambda}~\hat{\bf{e}}_{\textstyle{_{\lambda}}}~$.

 As we are employing the latitude, the right-handed triple changes to $\,\hat{\bf{e}}_{\textstyle{_{r}}}\,,\,\hat{\bf{e}}_{\textstyle{_{\lambda}}}\,,\,\hat{
 \bf{e}}_{\textstyle{_{\phi}}}\,$ (``radial -- east -- north"), and the gradient becomes $~\frac{\textstyle\partial U(\erbold)}{\textstyle\partial
 \erbold}~=~\frac{\textstyle\partial U}{\textstyle\partial r}~\hat{\bf{e}}_{\textstyle{_{r}}}\,+~\frac{\textstyle 1}{\textstyle r}~\frac{\textstyle 1}{\textstyle \cos\phi}~\frac{
 \textstyle \partial U}{\textstyle \partial \lambda}~\hat{\bf{e}}_{\textstyle{_{\lambda}}}\,-~\frac{\textstyle 1}{\textstyle r}~
 \frac{\textstyle\partial
 U}{\textstyle\partial \phi}~\hat{\bf{e}}_{\textstyle{_{\phi}}}~$.} Insertion of (\ref{therewith}) into (\ref{thereon}) results in
 \ba
 \vec{\bf T}~=~-~M_{sec}~\left(~0~\hat{\bf{e}}_{\textstyle{_{r}}}\,+~
 \frac{\partial U}{\partial \phi}~\hat{\bf{e}}_{\textstyle{_{\lambda}}}\,+~\frac{1}{\cos\phi}~\frac{\partial U}{\partial \lambda}~\hat{\bf{e}}_{\textstyle{_{\phi}}}~\right)~~,
 \label{du}
 \ea
 The torque $\,{\taubold}\,$ wherewith the secondary is acting on the primary will be the negative of (\ref{du}):
 \ba
 \taubold~=~M_{sec}~\left(~0~\hat{\bf{e}}_{\textstyle{_{r}}}\,+~
 \frac{\partial U}{\partial \phi}~\hat{\bf{e}}_{\textstyle{_{\lambda}}}\,+~\frac{1}{\cos\phi}~\frac{\partial U}{\partial \lambda}~\hat{\bf{e}}_{\textstyle{_{\phi}}}~\right)
 \label{torre}
 \ea
 Its east component, the one aimed along $\hat{\bf{e}}_{\textstyle{_{\lambda}}}$, is parallel to the primary's equatorial plane. The
 north component, aimed along $\hat{\bf{e}}_{\textstyle{_{\phi}}}$, is tangent to the meridian. In our paper however we employ the
 projection of the torque vector onto the primary's spin axis, i.e., the component $\,\tau_{\textstyle{_z}}\,$ orthogonal to the equator
 plane. This component is $\,\cos\phi\,$ times the north component:
  \ba
 {\cal{T}}_z~=~M_{sec}~\frac{\partial U}{\partial \lambda}~~~.
 \label{}
 \ea
 This torque component, equation (\ref{15}), slows down the rotation rate of the primary. The
 decelerating torque acting on the secondary has an opposite sign and is rendered by (\ref{T}).

 The east component and the projection of the north component onto the equator plane act to change the orientation of the primary's
 spin axis, a subject we shall not pursue in this paper.

 \section{Calculation of the average torque within the corrected MacDonald model}\label{averaging}

 To calculate the orbital average of the tidal torque acting on a librating secondary obeying the corrected MacDonald tidal model,
 substitute (\ref{qu}) and (\ref{duga}) into (\ref{19}), and then choose $\,\alpha=-1\,$ (and recall that, for $\,\alpha=-1\,$, the
 integral rheological parameter $\,{\cal{E}}\,$ is simply the time lag: $\,{\cal{E}}\,=\,\Delta t\,$). An equivalent option would be to
 employ (\ref{1919}) directly. This will lead us to the following expression for the torque:
 \ba
 \nonumber
 \left.~~~~~~~\right.{\cal{T}}_z ~= ~-~\frac{3}{2}~{GM_{sec}^2}\;k_{2}\,
 \frac{R^{\textstyle{^5}}}{r^{\textstyle{^{6}}}}\;{\cal{E}}\;\chi\;\,\mbox{sgn}(\dot{\theta}-
 \dot{\nu})+O(\inc^2/Q)+O(en\Delta t/Q)+O(Q^{-3})~~~~~~~~~~~~~~~~~~~~\\
  \nonumber\\
  \nonumber
 \left.\right.=~-~\frac{3}{2}~{GM_{sec}^2}\;k_{2} \frac{R^{\textstyle{^5}}}{r^{\textstyle{^{6}}}
 } \;2\;{\cal{E}}\;\,|\,\dot{\theta} - \dot{\nu}\,|\;\,\mbox{sgn}(\dot{\theta}-\dot{\nu})+
 O(\inc^2/Q)+O(en\Delta t/Q)+O(Q^{-3})~~~~~~~~~~\\
 \nonumber\\
 =\;-\;{3\;G\;M_{sec}^2}\;k_{2} \frac{R^{\textstyle{^5}}}{r^{\textstyle{^{6}}}
 }\;{\cal{E}}\;(\dot{\theta}-\dot{\nu})+O(\inc^2/Q)+O(en\Delta t/Q)+O(Q^{-3})
 ~~~,~~~~~~~~~~~~~~~~~~~~~~~~~~
 \label{toto}
 \ea
 and for its average over one orbiting cycle:
 \begin{subequations}
 \label{53}
 \ba
 \nonumber
 \langle\,{\cal{T}}_z\,\rangle=
 \,-\;\frac{3\,G\,M_{sec}^{\textstyle{^{\,2}}}\;\,k_{2}\;{\cal E}}{R}\;\;
 \langle\;\,(\dot{\theta}\,-\,\dot{\nu})\,\;
 \frac{R^{\textstyle{^6}}}{r^6}\;\;\rangle~~+O(\inc^2/Q)+O(Q^{-3})+O(en\Delta t/Q)
 \;=~~~
 ~~~~~~~~~~~~~~~~~~~~
 \label{}
 \ea
 \ba
 \left.~~~~~~~~\,\right.-\,\frac{3\,G\,M_{sec}^{\textstyle{^{\,2}}}\;k_{2}\,{\cal E}}{R}\;
 \dot{\theta}\;\;\langle\;\,\frac{R^{\textstyle{^6}}}{r^6}\;\;\rangle\,\;+\;\frac{3\,G\,
 M_{sec}^{\textstyle{^{\,2}}}\;\,k_{2}\;{\cal E}}{R}\;\langle\;\,\dot{\nu}\;
 \frac{R^{\textstyle{^6}}}{r^6}\;\;\rangle+O(\inc^2/Q)+O(Q^{-3})+O(en\Delta t/Q)~~~~~~\,
 \label{53a}
 \ea
 \ba
 \nonumber
  =&-&\frac{3\,G\,M_{sec}^{\textstyle{^{\,2}}}\;\,k_{2}\;{\cal E}}{R}\;\dot{\theta}\;\;
 \frac{R^6}{a^6}\left(1\,-\,e^2\right)^{-9/2}~\,\frac{1}{2\,\pi}\;
 \int_{0}^{2\pi}\left(1+e\;\cos\nu\right)^4\,d\nu\;~~~~
 ~~~~~~~~~~~~~~~~~~~~~\\
 \nonumber\\
 \nonumber\\
 &+&\frac{3\,G\,M_{sec}^{\textstyle{^{\,2}}}\;k_{2}\,{\cal E}}{R}\,n\,\frac{R^{\textstyle{^6}}}{a^6}
 \left(1-e^2\right)^{-6}\frac{1}{2\pi}\int_{0}^{2\pi}\left(1+e\,\cos\nu\right)^6
 d{\nu}+O(\inc^2/Q)+O(Q^{-3})+O(en\Delta t/Q)~~.~~~~~~~~~\,
 \label{53b}
 \ea
 \end{subequations}
 Evaluation of the integrals is straightforward and entails (\ref{Laskar} - \ref{54}).

 \end{document}